# Theory of the Anderson impurity model:
# The Schrieffer–Wolff transformation re–examined

Stefan K. Kehrein[1]  and Andreas Mielke[2]

Institut für Theoretische Physik,
Ruprecht–Karls–Universität,
D–69120 Heidelberg, F.R. Germany

October 5, 1995

**Abstract**

We apply the method of infinitesimal unitary transformations recently introduced by Wegner [1] to the Anderson single impurity model. It is demonstrated that this method provides a good approximation scheme for all values of the on-site interaction $U$, it becomes exact for $U = 0$. We are able to treat an arbitrary density of states, the only restriction being that the hybridization should not be the largest parameter in the system. Our approach constitutes a consistent framework to derive various results usually obtained by either perturbative renormalization in an expansion in the hybridization $\Gamma$, Anderson's "poor man's" scaling approach or the Schrieffer–Wolff unitary transformation. In contrast to the Schrieffer–Wolff result we find the correct high–energy cutoff and avoid singularities in the induced couplings. An important characteristic of our method as compared to the "poor man's" scaling approach is that we continuously decouple modes from the impurity that have a large energy difference from the impurity orbital energies. In the usual scaling approach this criterion is provided by the energy difference from the Fermi surface.

---

[1]E–mail: kehrein@marvin.tphys.uni-heidelberg.de
[2]E–mail: mielke@hybrid.tphys.uni-heidelberg.de

# 1 Introduction

In the past few years there has been renewed interest in the Anderson impurity model. This model was originally proposed by Anderson [2], for a recent review see [3]. It has been introduced to study the well–known Kondo problem, the behaviour of a single magnetic impurity coupled to a conduction band of electrons. The Hamiltonian contains the electron band, the energy of the impurity orbital together with a repulsive interaction on the impurity site, and a hybridization between the band states and the impurity state,

$$H = \sum_{k,\sigma} \epsilon_k c^\dagger_{k,\sigma} c_{k,\sigma} + \sum_\sigma \epsilon_d d^\dagger_\sigma d_\sigma + \sum_{k,\sigma} V_k (c^\dagger_{k,\sigma} d_\sigma + d^\dagger_\sigma c_{k,\sigma}) + U d^\dagger_+ d^\dagger_- d_- d_+. \tag{1.1}$$

In the case of a linear dispersion relation for the band and $V_k = V = $ const. the model was solved using a Bethe–ansatz [4]. But if one wants to study the Anderson impurity model in a more general situation, one needs a different approach. There are several methods available, most of them are reviewed in [3]. The most prominent method amongst them is probably the numerical renormalization group developed by Wilson [5] for the original Kondo Hamiltonian and applied to the Anderson impurity model by Krishna-murthy et al [6].

Recently we applied a new technique to this model [7]. We used continuous unitary transformations in a form introduced by Wegner [1] to diagonalize the Hamiltonian approximately. A continuous unitary transformation yields flow equations for the Hamiltonian, or equivalently flow equations for the coupling constants. The approximation we used neglected some additional couplings generated by the flow that were not present in the initial model. Unfortunately we were not able to obtain quantitative results. The reason is that the continuous transformations yields coupled, nonlinear differential equations for the different parameters in the Hamiltonian, which we were not able to treat analytically.

In a more recent work we applied continuous unitary transformation to the well known spin–boson model [8]. For this model a unitary transformation exists, the polaron transformation, which has been used to treat the Hamiltonian of the spin–boson model approximately. In Ref. [8] we used a simple modification of the previous ansatz by Wegner [1] for the generator of the continuous unitary transformation. Thereby we were able to construct a continuous polaron transformation. This new transformation does not have the disadvantages of the usual polaron transformation as it treats the slow bosonic modes in a satisfactory way. In addition, we were able to reduce the set of differential equations to a single, non–linear differential equation. This finally allowed us to obtain quantitative results which are in good agreement with results obtained by other methods.

In the case of the Anderson impurity model, a unitary transformation similar to the polaron transformation is known, the Schrieffer–Wolff transformation [9]. It has been introduced to eliminate the hybridization between the electronic states in the band and in the impurity orbital present in the Anderson model. Thereby it renormalizes the impurity energy and the repulsive interaction. Furthermore it generates a spin–spin interaction between the impurity electron and the conduction band electrons. This interaction is responsible for the Kondo effect. In this treatment one usually takes only the terms into account which are of second order in the hybridization. Other interactions are generated as well, but they are neglected.

The Schrieffer-Wolff transformation has some disadvantages. First it is equivalent to a second order perturbational treatment of the hybridization term in the Hamiltonian. Therefore the general validity of the result is unclear. In particular the Anderson impurity model in the Kondo regime is mapped onto a Kondo problem with an effective band width of order of the conduction band width. This is known to be wrong since the high–energy cutoff in the Anderson impurity model cannot be larger than of order the on-site interaction $U$ [10]. The second problem with the Schrieffer–Wolff transformation is that energy denominators occur, which become zero if the energy of the impurity level lies in the conduction band. If this is the case, the Schrieffer–Wolff transformation is ill–defined. This problem is similar to the problem of the treatment of the slow bosonic modes in the polaron transformation. Therefore we expect that a modification of our old treatment of the Anderson impurity model should be useful. In the present paper we construct a continuous, or infinitesimal Schrieffer–Wolff transformation in order to eliminate the hybridization terms in the Hamiltonian. We will show how such a continuous Schrieffer–Wolff transformation can be constructed systematically. Our old approach can be modified so that the number of additional interactions generated by the unitary transformation is reduced.

It is clear that applying infinitesimal unitary transformations to a given Hamiltonian is a non–perturbative method. The advantage of our present approach is that our method treats the Anderson impurity model consistently within one framework independent of whether the on–site interaction $U$ or the hybridization $\Gamma$ is the larger parameter. We are able to reduce the problem to two coupled non–linear differential equations for the impurity orbital energy and the on–site interaction. These are solved approximately yielding self–consistency conditions for these two quantities. Finally the antiferromagetic spin–spin interaction can be calculated. It is demonstrated how standard results from e.g. renormalization theory can be obtained in this conceptually new framework. Whereas other methods are not applicable in the whole parameter space or need additional assumptions, continuous unitary transformations are conceptually simple and no physically relevant restrictions or additional assumptions are needed. One can hope that the flow equations approach will be useful too for other problems with less well–established results. In so far it is important to study the method using a well–known problem.

Independent of Wegner [1], Glazek and Wilson [11, 12] have recently also proposed to use continuous unitary transformations to construct renormalization group equations for effective Hamiltonians in quantum field theory. Wilson et al. [13] applied this method to quantum chromodynamics. Although the general idea is similar, there are some differences between the approach of Glazek and Wilson and ours. Their goal is to eliminate "far–off–diagonal" matrix elements in a given Hamiltonian, which means off–diagonal matrix elements connecting states that are energetically far from each other. This means that the final Hamiltonian has a banded structure. In contrast our goal is to eliminate matrix elements such that the final Hamiltonian is diagonalized approximately or block–diagonal.

Although it is possible to construct flow equations such that the Hamiltonian becomes diagonal, it is not possible to solve these equations and to calculate the eigenenergies. Therefore we use the continuous unitary transformation in order to eliminate some of the matrix elements. In our case the hybridization is eliminated and the final Hamiltonian is block–diagonal. The final Hamiltonian does not contain matrix elements connecting states with a singly occupied impurity orbital to states for which the impurity orbital is either not occupied or doubly occupied. If now the impurity site is occupied with no or two electrons, the spin on the impurity site is zero and the additional antiferromagnetic interaction vanishes. In these cases the problem is essentially solved with respect to static properties. In the regime where the impurity site is singly occupied, the problem is reduced to a usual Kondo problem, which may be solved by various methods.

Our paper is organized as follows. In the following section we derive the general flow equations for the coupling constants of the Anderson impurity model. In section 3 we illustrate the method in the case of a vanishing interaction of the electrons on the impurity. It is shown that in this case the flow equations yield the exact solution. Furthermore we introduce an approximation which still gives the exact result for the case of vanishing interaction and which is applied in section 4 to the case of a non–vanishing interaction. We calculate implicit equations for the renormalized interaction and the renormalized impurity energy. These equations can in principle be solved for any given density of states and hybridization. Our results are compared with the results obtained by Schrieffer and Wolff [9]. In section 5 the antiferromagnetic spin–spin interaction is calculated. It can be used to determine e.g. the Kondo temperature. In section 6 we discuss the method of continuous unitary transformations as compared to a single unitary transformation in the Schrieffer–Wolff paper. Section 7 contains a discussion of our method as compared to the well–known "poor man's" scaling approach. The last section contains the conclusions and an outlook on other problems.

## 2 The flow equations for the Anderson impurity model

Our starting point is the Hamiltonian for the single impurity Anderson model (1.1), which we write in a normal ordered form

$$H = \sum_{k,\sigma} \epsilon_k : c_{k,\sigma}^\dagger c_{k,\sigma} : + \sum_{\sigma} \epsilon_d d_\sigma^\dagger d_\sigma + \sum_{k,\sigma} V_k (c_{k,\sigma}^\dagger d_\sigma + d_\sigma^\dagger c_{k,\sigma}) + U d_+^\dagger d_-^\dagger d_- d_+ + E_0. \tag{2.1}$$

We introduced a normal ordering for the band electron operators $c_{k,\sigma}^\dagger$ and $c_{k,\sigma}$. It is defined by $: c_{k,\sigma}^\dagger c_{k,\sigma} := c_{k,\sigma}^\dagger c_{k,\sigma} - n_k$ where $n_k = (\exp(\beta \epsilon_k) + 1)^{-1}$ is the occupation number of the band state with wave vector $k$. We let the Fermi energy equal to zero. The reason for the normal ordering is that additional interactions, which will be generated by our procedure and which we neglect, should be written down in a normal ordered form as well since otherwise the ground state expectation value of such additional contributions does not vanish. A detailed discussion has been given in [7]. In contrast to [7] we do not introduce a normal ordering for the impurity electron operators $d_\sigma^\dagger$ and $d_\sigma$. The reason is that in our approximation contributions containing such operators are not neglected if they have a non–vanishing expectation value in the ground state. It is possible to introduce a normal ordering on the impurity site as well, but the results are not changed. Due to the normal ordering a constant $E_0 = 2 \sum_k \epsilon_k n_k$ has been introduced in the Hamiltonian.

We now want to apply a general continuous unitary transformation to the Hamiltonian. Such a transformation is defined by a generator $\eta$ that depends on a continuous variable, which we call $\ell$. The continuous unitary transformation is defined by $\frac{dH}{d\ell} = [\eta(\ell), H(\ell)]$ with the initial condition $H(0) = H$ in (2.1). In order to simplify notation we will denote the initial values of $\epsilon_d$ and $U$ by

$$\epsilon_d^I \stackrel{\text{def}}{=} \epsilon_d(\ell = 0), \qquad U^I \stackrel{\text{def}}{=} U(\ell = 0) \tag{2.2}$$

and the asymptotic ("renormalized") values for $\ell \to \infty$ by

$$\epsilon_d^R \stackrel{\text{def}}{=} \epsilon_d(\ell = \infty), \qquad U^R \stackrel{\text{def}}{=} U(\ell = \infty). \tag{2.3}$$

If these parameters appear without an argument this will imply that they are to be considered as functions of $\ell$.

We choose $\eta$ to be of the form

$$\begin{aligned}
\eta &= \sum_{k,\sigma} \eta_k (c_{k,\sigma}^\dagger d_\sigma - d_\sigma^\dagger c_{k,\sigma}) + \sum_{k,q,\sigma} \eta_{k,q} (c_{k,\sigma}^\dagger c_{q,\sigma} - c_{q,\sigma}^\dagger c_{k,\sigma}) \\
&\quad + \sum_{k,\sigma} \eta_k^{(2)} (c_{k,\sigma}^\dagger d_{-\sigma}^\dagger d_{-\sigma} d_\sigma - d_\sigma^\dagger d_{-\sigma}^\dagger d_{-\sigma} c_{k,\sigma})
\end{aligned} \tag{2.4}$$

The commutator of $\eta$ and $H$ is easily calculated, we obtain

$$
\begin{aligned}
[\eta, H] &= \sum_{k,\sigma} \eta_k (\epsilon_d - \epsilon_k)(c^\dagger_{k,\sigma} d_\sigma + d^\dagger_\sigma c_{k,\sigma}) \\
&+ \sum_{k,q,\sigma} \eta_k V_q (: c^\dagger_{k,\sigma} c_{q,\sigma} : + : c^\dagger_{q,\sigma} c_{k,\sigma} :) \\
&- 2 \sum_{k,\sigma} \eta_k V_k d^\dagger_\sigma d_\sigma + 2 \sum_{k,\sigma} \eta_k V_k n_k \\
&+ U \sum_{k,\sigma} \eta_k (d^\dagger_\sigma d^\dagger_{-\sigma} d_{-\sigma} c_{k,\sigma} + c^\dagger_{k,\sigma} d^\dagger_{-\sigma} d_{-\sigma} d_\sigma) \\
&- \sum_{k,q,\sigma} \eta_{k,q} (\epsilon_k - \epsilon_q)(: c^\dagger_{k,\sigma} c_{q,\sigma} : + : c^\dagger_{q,\sigma} c_{k,\sigma} :) \\
&+ 2 \sum_{k,q,\sigma} \eta_{k,q} V_q (d^\dagger_\sigma c_{k,\sigma} + c^\dagger_{k,\sigma} d_\sigma) \\
&- \sum_{k,\sigma} \eta^{(2)}_k (\epsilon_k - \epsilon_d)(d^\dagger_\sigma d^\dagger_{-\sigma} d_{-\sigma} c_{k,\sigma} + c^\dagger_{k,\sigma} d^\dagger_{-\sigma} d_{-\sigma} d_\sigma) \\
&- 2 \sum_{k,\sigma} \eta^{(2)}_k V_k d^\dagger_\sigma d^\dagger_{-\sigma} d_{-\sigma} d_\sigma \\
&+ \sum_{k,q,\sigma} \eta^{(2)}_k V_q (: c^\dagger_{k,\sigma} d^\dagger_{-\sigma} d_{-\sigma} c_{q,\sigma} : + : c^\dagger_{q,\sigma} d^\dagger_{-\sigma} d_{-\sigma} c_{k,\sigma} : - : c^\dagger_{k,\sigma} d^\dagger_{-\sigma} d_\sigma c_{q,-\sigma} : \\
&\qquad - : c^\dagger_{q,\sigma} d^\dagger_{-\sigma} d_\sigma c_{k,-\sigma} : - c^\dagger_{k,\sigma} c^\dagger_{q,-\sigma} d_{-\sigma} d_\sigma - d^\dagger_\sigma d^\dagger_{-\sigma} c_{q,-\sigma} c_{k,\sigma}) \\
&+ 2 \sum_{k,\sigma} \eta^{(2)}_k V_k n_k d^\dagger_{-\sigma} d_{-\sigma} \\
&+ U \sum_{k,\sigma} \eta^{(2)}_k (d^\dagger_\sigma d^\dagger_{-\sigma} d_{-\sigma} c_{k,\sigma} + c^\dagger_{k,\sigma} d^\dagger_{-\sigma} d_{-\sigma} d_\sigma). \tag{2.5}
\end{aligned}
$$

Many additional couplings are generated which did not occur in the original Hamiltonian in (2.1). But some of these terms can be eliminated by a suitable choice of $\eta$. Let us first consider terms containing $: c^\dagger_{k,\sigma} c_{q,\sigma} + c^\dagger_{q,\sigma} c_{k,\sigma} :$. Such terms do not occur if we choose

$$\eta_{k,q}(\epsilon_k - \epsilon_q) = \frac{1}{2}(\eta_k V_q + \eta_q V_k). \tag{2.6}$$

Similarly, terms containing operators of the type $d^\dagger_\sigma d^\dagger_{-\sigma} d_{-\sigma} c_{k,\sigma} + c^\dagger_{k,\sigma} d^\dagger_{-\sigma} d_{-\sigma} d_\sigma$ do not occur if we choose

$$U\eta_k + (\epsilon_d - \epsilon_k)\eta^{(2)}_k + U\eta^{(2)}_k = 0 \tag{2.7}$$

These equations may be used to determine $\eta_{k,q}$ and $\eta^{(2)}_k$. The only contribution that does occur additionally is the term

$$
\begin{aligned}
\sum_{k,q,\sigma} \eta^{(2)}_k V_q (&: c^\dagger_{k,\sigma} d^\dagger_{-\sigma} d_{-\sigma} c_{q,\sigma} : + : c^\dagger_{q,\sigma} d^\dagger_{-\sigma} d_{-\sigma} c_{k,\sigma} : - : c^\dagger_{k,\sigma} d^\dagger_{-\sigma} d_\sigma c_{q,-\sigma} : \\
&- : c^\dagger_{q,\sigma} d^\dagger_{-\sigma} d_\sigma c_{k,-\sigma} : - c^\dagger_{k,\sigma} c^\dagger_{q,-\sigma} d_{-\sigma} d_\sigma - d^\dagger_\sigma d^\dagger_{-\sigma} c_{q,-\sigma} c_{k,\sigma}). \tag{2.8}
\end{aligned}
$$

Notice that the contributions containing $: c^\dagger_{k,\sigma} c_{q,\sigma} + c^\dagger_{q,\sigma} c_{k,\sigma} :$ or $: d^\dagger_\sigma d^\dagger_{-\sigma} d_{-\sigma} c_{k,\sigma} + c^\dagger_{k,\sigma} d^\dagger_{-\sigma} d_{-\sigma} d_\sigma :$ have been classified irrelevant in our former approach [7], whereas the term in (2.8) is marginal in some of the fixed points [7]. In principle it has to be included in the Hamiltonian in (2.1). But the commutator of this term with $\eta$ does not yield contributions to the other terms in the Hamiltonian. Therefore we do not take it into account in our first analysis of the problem. But it is clear that this additional term is important. A part of it yields the antiferromagnetic interaction between the impurity spin and the spins of the band electrons that is responsible for the Kondo effect. We will come back to this interaction later. Our first goal is to calculate the flow equations for the parameters in the Hamiltonian (2.1). From (2.5) we obtain

$$\frac{dV_k}{d\ell} = \eta_k(\epsilon_d - \epsilon_k) + 2\sum_p \eta_{k,p} V_p \tag{2.9}$$

$$\frac{d\epsilon_k}{d\ell} = 2\eta_k V_k \tag{2.10}$$

$$\frac{d\epsilon_d}{d\ell} = -2\sum_k \eta_k V_k + 2\sum_k \eta_k^{(2)} V_k n_k \tag{2.11}$$

$$\frac{dU}{d\ell} = -4\sum_k \eta_k^{(2)} V_k \tag{2.12}$$

$$\frac{dE_0}{d\ell} = 4\sum_k \eta_k V_k n_k \tag{2.13}$$

The last equation yields directly $E_0 = 2\sum_k \epsilon_k n_k$, which is the energy of the filled Fermi sea. In the following we are interested in the thermodynamic limit. For large $N$, the number of states in the band, one has $V_k \propto N^{-1/2}$. Thus, $\eta_k$ must as well be of the order $N^{-1/2}$, and the derivative of $\epsilon_k$ with respect to $\ell$ is of order $N^{-1}$. For large values of $N$ the band energies do not depend on $\ell$. This should have been expected. The thermodynamic bath of electrons is not affected by the single impurity. This means that the global density of states in the band is fixed. But this does not mean that the local density of states is fixed as well. In contrary, one should expect that the local density of states near the impurity site is affected by the impurity. We will come back to this point in the discussion.

## 3 Vanishing interaction $U = 0$

To illustrate the advantages of our method, let us first study the case $U = 0$. Then we have a quadratic Hamiltonian that can be solved exactly, see for example Ref. [14]. We will show that our method yields the exact solution in this case. For $U = 0$ the flow equations simplify to

$$\frac{dV_k}{d\ell} = \eta_k(\epsilon_d - \epsilon_k) + 2\sum_p \eta_{k,p} V_p \tag{3.1}$$

$$\frac{d\epsilon_d}{d\ell} = -2\sum_k \eta_k V_k \tag{3.2}$$

The flow equations are exact in the case $U = 0$ since the neglected terms in (2.5) vanish in this limit. We let

$$\eta_k = V_k f(\epsilon_k, \ell) \tag{3.3}$$

and introduce

$$J(\epsilon, \ell) = \sum_k V_k^2 \delta(\epsilon - \epsilon_k). \tag{3.4}$$

In the literature one often introduces the parameter

$$\Gamma = \pi \rho(\epsilon_F) V_{k_F}(0)^2 = \pi J(\epsilon_F, 0), \quad (3.5)$$

where $\rho(\epsilon_F)$ is the density of states at the Fermi surface. The flow equations for $\epsilon_d$ and $J(\epsilon, \ell)$ are

$$\frac{d\epsilon_d}{d\ell} = -2 \int d\epsilon f(\epsilon, \ell) J(\epsilon, \ell) \quad (3.6)$$

$$\frac{\partial J(\epsilon, \ell)}{\partial \ell} = 2 J(\epsilon, \ell) f(\epsilon, \ell)(\epsilon_d - \epsilon) + 2 \int d\epsilon' \frac{J(\epsilon, \ell) J(\epsilon', \ell)(f(\epsilon, \ell) + f(\epsilon', \ell))}{\epsilon - \epsilon'}. \quad (3.7)$$

For the integral in the last equation one has to take its principal value. The set of flow equations may be solved if one introduces a function

$$G(\epsilon, \ell) = \int d\epsilon' \frac{J(\epsilon', \ell)}{\epsilon - \epsilon' + G(\epsilon, \ell)}. \quad (3.8)$$

Taking the derivative with respect to $\ell$, we obtain an implicit equation for this derivative, which can be solved. The final result is

$$\frac{\partial G(\epsilon, \ell)}{\partial \ell} = 2 \left[ 1 + \int d\epsilon' \frac{J(\epsilon', \ell)}{(\epsilon - \epsilon' + G(\epsilon, \ell))^2} \right]^{-1} \int d\epsilon' f(\epsilon', \ell) J(\epsilon', \ell) \frac{\epsilon_d - \epsilon' + G(\epsilon, \ell)}{\epsilon - \epsilon' + G(\epsilon, \ell)} \quad (3.9)$$

Calculating the derivative of $G(\epsilon, \ell)$ with respect to $\epsilon$ we obtain similarly

$$1 + \frac{\partial G(\epsilon, \ell)}{\partial \epsilon} = \left[ 1 + \int d\epsilon' \frac{J(\epsilon', \ell)}{(\epsilon - \epsilon' + G(\epsilon, \ell))^2} \right]^{-1}. \quad (3.10)$$

This yields

$$\frac{\frac{\partial G(\epsilon, \ell)}{\partial \ell}}{1 + \frac{\partial G(\epsilon, \ell)}{\partial \epsilon}} = 2 \int d\epsilon' f(\epsilon', \ell) J(\epsilon', \ell) \frac{\epsilon_d - \epsilon' + G(\epsilon, \ell)}{\epsilon - \epsilon' + G(\epsilon, \ell)} \quad (3.11)$$

Comparing the right hand side with the derivative of $\epsilon_d$ with respect to $\ell$, we obtain

$$\frac{d\epsilon_d}{d\ell} = - \left. \frac{\frac{\partial G(\epsilon, \ell)}{\partial \ell}}{1 + \frac{\partial G(\epsilon, \ell)}{\partial \epsilon}} \right|_{\epsilon = \epsilon_d(\ell)}. \quad (3.12)$$

This equation can be integrated and the final result is

$$\epsilon_d(\ell) + G(\epsilon_d(\ell), \ell) = \epsilon_d^R. \quad (3.13)$$

In the last step we used $G(\epsilon, \infty) = 0$, which follows directly from $J(\epsilon, \infty) = 0$ and holds for an appropriate choice of $f(\epsilon, \ell)$. Solving for $\epsilon_d(\ell)$, we obtain

$$\epsilon_d(\ell) = \epsilon_d^R - \int d\epsilon \frac{J(\epsilon, \ell)}{\epsilon_d^R - \epsilon} \quad (3.14)$$

Again, for the integral on the right hand side we have to take its principal value. This means that we have to choose $J(\epsilon, \ell)$ and therefore $f(\epsilon, \ell)$ such that the principal value exists for all $\ell$. We obtain the value of $\epsilon_d^R$, if we let $\ell = 0$ and solve for $\epsilon_d^R$.

As a simple example we take a semi-circle

$$J(\epsilon, 0) = \frac{2V^2}{\pi D^2} \sqrt{D^2 - \epsilon^2} \quad (3.15)$$

where $2D$ is the band width and $V = \sqrt{\sum_k V_k^2}$. Here we have $\Gamma = 2V^2/D$. The main reason for this choice of the hybridization is that all the integrals can be worked out in closed form in the sequel. But it should be noted that it is a main advantage of our approach that it can be used for arbitrary functions $J(\epsilon, 0)$, in particular for any distribution of the density of states in the conduction band. However, if one chooses a linear dispersion relation and constant hybridization $V_k$, that is $J(\epsilon, 0) = \frac{V^2}{2D}\Theta(D - |\epsilon|)$ as usually done in renormalization group treatment of the Anderson impurity model, one must be careful due to the discontinuous behaviour close to the band edge. The self–consistency equations in the flow equations approach will then generally have more than one solution, however, the actual solution of the differential equations chooses the correct one. Close to the band edge one expects unphysical behaviour anyway due to the unphysical choice of $J(\epsilon, 0)$ and neither approach should be trusted. Therefore it is natural in our approach to choose a function $J(\epsilon, 0)$ that is continuous at the band edge as should be expected on physical grounds anyway.

Let us come back to our example introduced in Eq. (3.15). We have to distinguish between the case where $\epsilon_d^R$ lies in the band and the case where it lies outside the band. We first consider the latter case. The integral can be calculated and we obtain

$$\epsilon_d^R - \epsilon_d^I + \frac{\Gamma}{D}\left(\text{sign}(\epsilon_d^R)\sqrt{(\epsilon_d^R)^2 - D^2} - \epsilon_d^R\right) = 0 \tag{3.16}$$

if $\left|\epsilon_d^R\right| > D$. This equation yields a simple quadratic equation for $\epsilon_d^R$, which has always two solutions. If $\Gamma < D$ at most one of these solutions lies outside the band. If $\Gamma > D$ and $\left|\epsilon_d^I\right| > \Gamma - D$, there is a single solution for $\epsilon_d^R$ outside the band, but if $\left|\epsilon_d^I\right| < \Gamma - D$ we obtain two solutions outside the conduction band.

The situation is much simpler when $\epsilon_d^R$ lies inside the band. The integral in (3.14) has to be interpreted as its principal value and we obtain

$$\epsilon_d^R - \epsilon_d^I - \frac{\Gamma}{D}\epsilon_d^R = 0. \tag{3.17}$$

The only solution is

$$\epsilon_d^R = \frac{\epsilon_d^I}{1 - \frac{\Gamma}{D}}. \tag{3.18}$$

$\epsilon_d^R$ lies inside the band if $\Gamma < D - \left|\epsilon_d^I\right|$. The various cases are shown in Fig. 1.

The fact that for a sufficiently large value of $V$ ($\Gamma > D$ in our example) two solutions for $\epsilon_d^R$ exist, is generic. It holds for any $J(\epsilon, 0)$ with a connected support of length $2D$. It is clear that with the present approach of flow equations only a single solution can be obtained. Nevertheless, the second solution is of physical importance. It is possible that a localized state develops from the original band states that has an energy which lies outside the band. Such a state cannot be obtained within the present formulation of the flow equations. In the case $U = 0$ one can introduce a different representation of the Hamiltonian and of $\eta$ that includes a localized band state explicitly. Since we are at present not able to deal with similar problems in the case $U > 0$, we restrict ourselves to the parameter regime $\Gamma < D$ (in fact later we will need $\Gamma < D/2$). This is reasonable from a physical point of view since we do not expect that the hybridization is the largest parameter in the system.

Eq. (3.14) is obtained as well if one uses two simple approximations to the flow equation. The first approximation neglects the terms proportional to $c_{k,\sigma}^\dagger c_{q,\sigma}$ in the Hamiltonian that are generated by the transformation and consequently one neglects such terms in $\eta$ as well. This is an approximation that can be justified from a physical point of view, since these terms are irrelevant in all fixed points [7]. Then the second term in (2.9) vanishes and the equation for $V_k$ is linear in $V_k$. Similarly, the second term in (3.7) vanishes, whereas (3.6) remains unchanged. Both equations together yield

$$\frac{d\epsilon_d}{d\ell} = -\int d\epsilon \frac{\frac{\partial J(\epsilon,\ell)}{\partial \ell}}{\epsilon_d - \epsilon}. \tag{3.19}$$

Furthermore we assume that $\epsilon_d(\ell)$ converges rapidly to $\epsilon_d^R$, so that we can replace $\epsilon_d(\ell)$ with $\epsilon_d^R$ on the right hand side (3.19). This yields (3.14). We will use similar approximations for $U > 0$ as well. Although it is possible to choose $\eta$ in such a way that only very few new terms are generated, the flow equations become very complicated. In order to be able to analyse the flow equations, one has to neglect higher interactions. This is often possible due to physical reasons.

One would like to understand why the self–consistency condition obtained by replacing $\epsilon_d$ by $\epsilon_d^R$ on the right hand side of (3.19) yields a good approximation to the exact solution of (3.19). To discuss this point let us introduce a special choice of $f(\epsilon, \ell)$. Since we want $J(\epsilon, \ell)$ to vanish in the limit $\ell \to \infty$, a natural choice would be $f(\epsilon, \ell) = -(\epsilon_d - \epsilon)$. But for finite $U$ we will have to make a different choice for $f(\epsilon, \ell)$ in the next section. For consistency we therefore take $f(\epsilon, \ell) = -(\epsilon_d - \epsilon)^3/(4\epsilon_d^2)$. This obviously works as well and it is easy to see that in the present case both choices are essentially equivalent.

The following argument holds in both cases. Unless $\epsilon = \epsilon_d$, $J(\epsilon, \ell)$ decays exponentially on a scale set by $\ell \propto \epsilon_d^2/(\epsilon_d - \epsilon)^4$. If $\epsilon_d$ lies outside the band it will tend to $\epsilon_d^R$ exponentially and the approximation $\epsilon_d \approx \epsilon_d^R$ on the right hand side of (3.19) is justified. On the other hand, if $\epsilon_d$ lies inside the band, we can estimate the relevant $\ell$–scale on which $\epsilon_d$ changes by calculating the ratio of the total change of $\epsilon_d$ to its derivative with respect to $\ell$ for small $\ell$. This shows that $\epsilon_d$ changes on a scale set by $\ell \propto \epsilon_d^2/D^4$, i.e. much faster than $J(\epsilon, \ell)$ for values of $\epsilon$ near the Fermi energy. Therefore $\epsilon_d$ can be replaced by its renormalized value on the right hand side of (3.19). We will use the same approximation in the next section to discuss the case $U > 0$, it can be justified in the same manner.

## 4 Non–vanishing interaction $U > 0$

With the approximations introduced at the end of the last section, the flow equations for $U > 0$ may be written in the form

$$\frac{dV_k}{d\ell} = \eta_k(\epsilon_d - \epsilon_k), \tag{4.1}$$

$$\frac{d\epsilon_d}{d\ell} = -2\sum_k \eta_k V_k + 2\sum_k \eta_k^{(2)} V_k n_k, \tag{4.2}$$

$$\frac{dU}{d\ell} = -4\sum_k \eta_k^{(2)} V_k. \tag{4.3}$$

According to (2.7) we take $\eta_k^{(2)} = -U\eta_k(\epsilon_d - \epsilon_k + U)^{-1}$ and as above $\eta_k = V_k f(\epsilon_k, \ell)$. We assume that $\sum_k V_k^2 < D^2$ so that the renormalized $\epsilon_d^R$ is unique for $U = 0$. We expect that it is unique for $U > 0$ as well. With these assumptions we proceed as in the previous section. We introduce $J(\epsilon, \ell)$ as in (3.4) and obtain the flow equations

$$\frac{\partial J(\epsilon, \ell)}{\partial \ell} = 2f(\epsilon, \ell)J(\epsilon, \ell)(\epsilon_d - \epsilon), \tag{4.4}$$

$$\frac{d\epsilon_d}{d\ell} = -\int d\epsilon \frac{\partial J(\epsilon,\ell)}{\partial \ell} \frac{\epsilon_d - \epsilon + (1+n(\epsilon))U}{(\epsilon_d - \epsilon)(\epsilon_d - \epsilon + U)}, \tag{4.5}$$

$$\frac{dU}{d\ell} = 2\int d\epsilon \frac{\partial J(\epsilon,\ell)}{\partial \ell} \frac{U}{(\epsilon_d - \epsilon)(\epsilon_d - \epsilon + U)}. \tag{4.6}$$

$n(\epsilon)$ is the Fermi distribution. The first equation may be used to parametrize $J(\epsilon,\ell)$. A suitable parametrization is

$$J(\epsilon,\ell) = J(\epsilon,0)\exp\left(-\int_0^\ell \frac{(\epsilon_d - \epsilon)^2(\epsilon_d - \epsilon + U)^2}{\epsilon_d^2 + (\epsilon_d + U)^2} d\ell'\right). \tag{4.7}$$

We will see that with this choice the hybridization flows to zero for all $\epsilon$, in particular also for $\epsilon = \epsilon_d^R$ or $\epsilon = \epsilon_d^R + U^R$. The reason is that $\epsilon_d - \epsilon_d^R$ decays like $\ell^{-1/2}$ as we will see below. $J(\epsilon_d^R, \ell)$ decays algebraically to zero. Furthermore, Eq. (4.7) corresponds to the following function $f(\epsilon,\ell)$

$$f(\epsilon,\ell) = -\frac{(\epsilon_d - \epsilon)(\epsilon_d - \epsilon + U)^2}{2\epsilon_d^2 + 2(\epsilon_d + U)^2}. \tag{4.8}$$

The reason for this choice of $f(\epsilon,\ell)$ or $J(\epsilon,\ell)$ is that now no pole terms appear in the integrals on the right hand sides of (4.5) and (4.6). The denominator in (4.8) is just introduced for convenience so that limits like $\lim_{U\to\infty}$ can be performed in all the equations without difficulties. Later we will come back to the question of other parametrizations $f(\epsilon,\ell)$. In fact Eq. (4.8) belongs to a class of parametrizations that all give the same physical results, whereas other parametrizations lead to divergencies or $J(\epsilon,\ell)$ does not flow to zero everywhere.

Let us consider for a moment the simplified case $\lim_{U\to\infty}$. The equation for $\epsilon_d$ takes the form

$$\frac{d\epsilon_d}{d\ell} = \int d\epsilon J(\epsilon,0)(1+n(\epsilon))(\epsilon_d - \epsilon)\exp\left(-\int_0^\ell (\epsilon_d - \epsilon)^2 d\ell'\right). \tag{4.9}$$

This equation is very similar to the flow equation for the renormalized tunneling frequency in the spin-boson problem [8]. The asymptotic behaviour can be obtained as in [8], one finds $\epsilon_d^R - \epsilon_d(\ell) \propto \ell^{-\frac{1}{2}}$ for large $\ell$ if $\epsilon_d$ lies inside the band. Otherwise it decays exponentially. (4.9) shows that $1/\sqrt{\ell}$ plays the role of an effective band width if $1/\sqrt{\ell}$ becomes smaller than the original band width of $J(\epsilon,0)$. The effective band width is reduced with increasing $\ell$. This is similar to a renormalization group procedure, where modes with high energies are integrated out. In our case, these modes are decoupled from the system. The analogies with renormalization theory will be worked out in more detail in section 7.

For finite $U$ the situation is somewhat more complicated, but the results are similar. The effective band width is $1/\sqrt{\ell}$ and leads to an asymptotic behaviour $U^R - U(\ell) = C_1 \ell^{-\frac{1}{2}}$ and $\epsilon_d^R - \epsilon_d(\ell) = -C_2 \ell^{-\frac{1}{2}}$ for large $\ell$ with some constants $C_1$ and $C_2$. This again holds if $\epsilon_d$ and $\epsilon_d + U$ lie inside the band. $C_1$ and $C_2$ are positive if $\epsilon_d$ lies below and $\epsilon_d + U$ lies above the Fermi energy. One possibility to obtain these results is to make the ansatz $U(\ell) = U^R + C_1 \ell^\alpha$ and $\epsilon_d(\ell) = \epsilon_d^R - C_2 \ell^\beta$. Inserting these expressions in the flow equations one shows easily that $\alpha = \beta = \frac{1}{2}$ is the only possible solution. We now replace $U$ and $\epsilon_d$ by their asymptotic values on the right hand side of (4.5) and (4.6). Both equations can be integrated and we obtain

$$\epsilon_d(\ell) = \epsilon_d^R - \int d\epsilon J(\epsilon,\ell) \frac{\epsilon_d^R - \epsilon + (1+n(\epsilon))U^R}{(\epsilon_d^R - \epsilon)(\epsilon_d^R - \epsilon + U^R)}, \tag{4.10}$$

$$U(\ell) = U^R + 2\int d\epsilon J(\epsilon,\ell) \frac{U^R}{(\epsilon_d^R - \epsilon)(\epsilon_d^R - \epsilon + U^R)}. \tag{4.11}$$

These equations are good approximations to the solution of the flow equations (4.5) and (4.6). They give the correct asymptotic behaviour and a numerical integration of the flow equations shows that the true solution differs only slightly from the approximate value. It should be noted that this is a different level of approximation than the previous restriction to a certain set of interactions included in the flow equations. This restriction was a physical approximation whereas the approximate solutions (4.10) and (4.11) can be controlled by solving the original differential equations numerically. We have done that too and always found very good agreement.

For $U = 0$ we showed that (4.10) yields the exact result. For $U = \infty$, (4.10) is correct up to terms quadratic in $J(\epsilon, \ell)$. This can be seen if one notices that (4.10) is the exact solution of a set of equations similar to (3.6) and (3.7) but with $J(\epsilon, \ell)$ replaced by $J(\epsilon, \ell)(1 + n(\epsilon))$. An additional argument to justify this approach is similar to the one given at the end of the previous section. The relevant $\ell$–scale for changes of $\epsilon_d$ and $U$ is smaller than the scale on which $J(\epsilon, \ell)$ varies. The crossover to the asymptotic behaviour occurs for $\ell > D^{-2}$, whereas $J(\epsilon, \ell)$ does not change too much on this scale. Taking $\ell = 0$, (4.10) and (4.11) yield the self–consistency conditions for $\epsilon_d^R$ and $U^R$. Let us mention that the results for $\epsilon_d^R$ and $U^R$ obtained from (4.10) and (4.11) do not depend on the special choice of $f(\epsilon, \ell)$ in (4.8). Nevertheless (4.10) and (4.11) are only good approximations to the flow equations (4.5) and (4.6) for special choices of $f(\epsilon, \ell)$ like the one in (4.8). The important point is that $J(\epsilon, \ell)$ has to be chosen so that the principal value of the integrals in (4.10) and (4.11) is well–defined for all values of $\ell$. This is clearly true for $J(\epsilon, \ell)$ given in (4.7). Our results here do not depend on the details of the continuous unitary transformation. But the transformation has to be chosen such that the flow for all the parameters in the Hamiltonian is well–defined. In section 7 we will see an example of a different parametrization of $J(\epsilon, \ell)$ where this is not the case.

Let us again consider the case $J(\epsilon, 0) = \frac{2V^2}{\pi D^2}\sqrt{D^2 - \epsilon^2}$. The equation for $U^R$ does not contain a factor $n(\epsilon)$ and the integral is easily evaluated. The result is

$$U^I = U^R - \frac{2\Gamma}{D}\left[U^R - \theta(\left|\epsilon_d^R + U^R\right| - D)\mathrm{sign}(\epsilon_d^R + U^R)\sqrt{(\epsilon_d^R + U^R)^2 - D^2}\right.$$
$$\left. + \theta(\left|\epsilon_d^R\right| - D)\mathrm{sign}(\epsilon_d^R)\sqrt{(\epsilon_d^R)^2 - D^2}\right]. \tag{4.12}$$

This equation shows that if $\epsilon_d^R$ lies below the Fermi energy and $\epsilon_d^R + U^R$ lies above the Fermi energy, then $U^R$ is larger than the initial value $U^I$. This is also true if both, $\epsilon_d^R$ and $\epsilon_d^R + U^R$ lie in the band. It this case we simply obtain

$$U^R = \frac{U^I}{1 - \frac{2\Gamma}{D}}. \tag{4.13}$$

On the other hand, if $\epsilon_d^R$ and $\epsilon_d^R + U^R$ lie above the energy band, $U^R$ is smaller than $U^I$.

**Symmetric Anderson model**

In the symmetric case $U^I = -2\epsilon_d^I$, the flow equations yield $U(\ell) = -2\epsilon_d(\ell)$ and the above conditions give $U^R = -2\epsilon_d^R$ as it should be. Therefore we have

$$\epsilon_d^I = \epsilon_d^R + \frac{2\Gamma}{D}\left[\theta(\left|\epsilon_d^R\right| - D)\mathrm{sign}(\epsilon_d^R)\sqrt{(\epsilon_d^R)^2 - D^2} - \epsilon_d^R\right]. \tag{4.14}$$

This equation is very similar to the one obtained for $U = 0$. If $\left|\epsilon_d^I\right| < D - 2\Gamma$ the renormalized value is $\epsilon_d^R = \epsilon_d^I/(1 - 2\Gamma/D)$ and lies inside the band. If $\left|\epsilon_d^I\right| > D - 2\Gamma$ we obtain a quadratic equation for $\epsilon_d^R$. For $2\Gamma < D$ this equation has a single solution outside the band, whereas for $2\Gamma > D$ we can obtain two solutions of the self–consistency equations outside the conduction band similar to the case $U = 0$.

**Asymmetric Anderson model**

In the general case ($U \neq -2\epsilon_d$), we have to calculate the integral in (4.10). It contains a factor $n(\epsilon)$ due to the normal ordering we introduced in the Hamiltonian. Therefore the renormalized impurity energy $\epsilon_d^R$ depends on the temperature and the chemical potential. We let $T = 0$ and $\epsilon_F = 0$, so that $n(\epsilon) = 1 - \theta(\epsilon)$. In this case the integrals can easily be evaluated explicitly. We have to distinguish various cases of whether the impurity orbital energies lie inside the conduction band or outside.

- $\left|\epsilon_d^R\right|, \left|\epsilon_d^R + U^R\right| > D$:
  We obtain

$$\epsilon_d^I = \epsilon_d^R - \frac{\Gamma}{2D}\left[2\epsilon_d^R - U^R + \text{sign}(\epsilon_d^R + U^R)\sqrt{(\epsilon_d^R + U^R)^2 - D^2}\left(1 - \frac{2}{\pi}\arcsin\frac{D}{\epsilon_d^R + U^R}\right)\right.$$
$$\left. - \text{sign}(\epsilon_d^R)\sqrt{(\epsilon_d^R)^2 - D^2}\left(3 - \frac{2}{\pi}\arcsin\frac{D}{\epsilon_d^R}\right)\right]. \quad (4.15)$$

In the limit $U^I = 0$ we have $U^R = 0$ and the condition for $\epsilon_d^R$ is the same as in the previous section. In the limit $U = \infty$, we obtain a single equation for $\epsilon_d^R$,

$$\epsilon_d^I = \epsilon_d^R + \frac{\Gamma}{2D}\left[\text{sign}(\epsilon_d^R)\sqrt{(\epsilon_d^R)^2 - D^2}\left(3 - \frac{2}{\pi}\arcsin\frac{D}{\epsilon_d^R}\right) + \frac{2}{\pi}D - 3\epsilon_d^R\right]. \quad (4.16)$$

In both expressions, $\left|\epsilon_d^R\right| > \left|\epsilon_d^I\right|$. The impurity orbital energy is pushed away from the band as it should have been expected.

- $\left|\epsilon_d^R\right| < D, \left|\epsilon_d^R + U^R\right| > D$:
  We obtain

$$\epsilon_d^I = \epsilon_d^R - \frac{\Gamma}{2D}\left[2\epsilon_d^R - U^R + \frac{2}{\pi}\sqrt{D^2 - (\epsilon_d^R)^2}\ln\left(\frac{\sqrt{D^2 - (\epsilon_d^R)^2} + D}{|\epsilon_d^R|}\right)\right.$$
$$\left. + \text{sign}(\epsilon_d^R + U^R)\sqrt{(\epsilon_d^R + U^R)^2 - D^2}\left(1 - \frac{2}{\pi}\arcsin\frac{D}{\epsilon_d^R + U^R}\right)\right]. \quad (4.17)$$

If we let $U = \infty$ in this case, this expression simplifies to

$$\epsilon_d^I = \epsilon_d^R - \frac{\Gamma}{2D}\left[3\epsilon_d^R - \frac{2}{\pi}D + \frac{2}{\pi}\sqrt{D^2 - (\epsilon_d^R)^2}\ln\left(\frac{\sqrt{D^2 - (\epsilon_d^R)^2} + D}{|\epsilon_d^R|}\right)\right]. \quad (4.18)$$

Finally for $\left|\epsilon_d^R\right| \ll D \ll U$

$$\epsilon_d^R = \epsilon_d^I + \frac{\Gamma}{\pi} \ln\left|\frac{2D}{e\,\epsilon_d^R}\right| + \Gamma\, O(\frac{D}{U^R}). \tag{4.19}$$

This is shown in Fig. 2. Eq. (4.19) contains a logarithmic singularity on the right hand side for $\epsilon_d^R \to 0$. If $\epsilon_d^I$ is negative and sufficiently far away from the Fermi energy, $\epsilon_d^R$ is negative as well and increases with increasing $\epsilon_d^I$. At some (still negative) value of $\epsilon_d^I$ the renormalized impurity energy jumps discontinuously from a given value below the Fermi energy to a value above the Fermi energy and then increases further with increasing $\epsilon_d^I$. This behaviour can be deduced from the numerical solution of the differential equations and is depicted by the full line in Fig. 2. However, it is difficult to obtain reliable numerical results in this regime. $\epsilon_d^R$ never reaches the Fermi level except for the trivial case where $V=0$ and $\epsilon_d^I = 0$.

- $\left|\epsilon_d^R\right|, \left|\epsilon_d^R + U^R\right| < D$:

We obtain

$$\begin{aligned}\epsilon_d^I &= \epsilon_d^R - \frac{\Gamma}{2D}\left[2\epsilon_d^R - U^R + \frac{2}{\pi}\sqrt{D^2 - (\epsilon_d^R)^2}\ln\left(\frac{\sqrt{D^2 - (\epsilon_d^R)^2} + D}{|\epsilon_d^R|}\right)\right.\\ &\quad \left. - \frac{2}{\pi}\sqrt{D^2 - (\epsilon_d^R + U^R)^2}\ln\left(\frac{\sqrt{D^2 - (\epsilon_d^R + U^R)^2} + D}{|\epsilon_d^R + U^R|}\right)\right].\end{aligned} \tag{4.20}$$

This expression contains two logarithmic singularities, it diverges if either $\epsilon_d^R \to 0$ or $\epsilon_d^R + U^R \to 0$. As a consequence, both $\epsilon_d^R$ and $\epsilon_d^R + U^R$ cannot approach the Fermi energy as long as $V \neq 0$. For $\left|\epsilon_d^R\right| \ll U^R \ll D$ one finds in particular

$$\epsilon_d^R = \epsilon_d^I + \frac{\Gamma}{\pi}\ln\left|\frac{U^R}{\epsilon_d^R}\right| + \Gamma\, O(\frac{U^R}{D}). \tag{4.21}$$

Again the solution of Eq. (4.21) is non–unique for some range of the initial parameter $\epsilon_d^I$. In fact Eq. (4.21) is well–known from renormalization theory [6, 10, 16]: In the valence fluctuation regime of the asymmetric Anderson model one has to replace $\epsilon_d^I$ by an *effective impurity orbital energy* $E_d^*$ (we use the notation from [6]). $E_d^*$ can be obtained as the solution of the following equation

$$T_2^* = -E_d(T_2^*) \stackrel{\text{def}}{=} -E_d^* \tag{4.22}$$

with

$$-E_d(T) = -\epsilon_d^I - \frac{\Gamma}{\pi}\ln\left(\frac{U}{T}\right). \tag{4.23}$$

One easily checks $E_d^* = \epsilon_d^R$ with $\epsilon_d^R$ from the flow equations approach. This result for $\epsilon_d^R$ will play an important role for the calculation of the Kondo temperature in this regime in Sect. 5. It will then become apparent why it is important to find the effective impurity orbital energy from renormalization theory to be identical to our asymptotic value $\epsilon_d^R$.

Finally it should be emphasized that the flow equations immediately give the correct high–energy cutoff in Eqs. (4.19) and (4.21). The smaller of the two parameters $U$ and $D$ appears in the logarithm and this comes about here very naturally.

These results, especially the self–consistency conditions in the general form

$$\epsilon_d^R = \epsilon_d^I + \int d\epsilon J(\epsilon, 0) \frac{\epsilon_d^R - \epsilon + (1 + n(\epsilon))U^R}{(\epsilon_d^R - \epsilon)(\epsilon_d^R - \epsilon + U^R)} \tag{4.24}$$

$$U^R = U^I - 2 \int d\epsilon J(\epsilon, 0) \frac{U^R}{(\epsilon_d^R - \epsilon)(\epsilon_d^R - \epsilon + U^R)} \tag{4.25}$$

can be compared with the result of the Schrieffer–Wolff transformation. If one applies a usual Schrieffer–Wolff transformation to the Hamiltonian, the impurity energy $\epsilon_d$ and the interaction $U$ are renormalized as well. But the result is a simple result of a second order perturbational treatment. It may be obtained from our self–consistency conditions if they are solved recursively to first order in $J(\epsilon, 0)$. This simply means that on the right hand sides of (4.24) and (4.25) the renormalized values are approximated by the initial values. As already mentioned our result is exact if $U = 0$. In this case the self–consistency condition for $\epsilon_d^R$ can be solved iteratively. This corresponds to summing up the whole perturbational series. Similarly, the self–consistency conditions (4.24) and (4.25) can be solved recursively though this will in general not give the exact result. But it is seems that a large part of the perturbational series is summed up when one solves these equations due to the same reasons that were already mentioned at the end of section 3.

One of the main results that we have obtained for $J(\epsilon, 0) = \frac{2V^2}{\pi D^2}\sqrt{D^2 - \epsilon^2}$ was that the renormalized values $\epsilon_d^R$ and $\epsilon_d^R + U^R$ behave discontinuously at the Fermi energy as a function of the initial values $\epsilon_d^I$ and $\epsilon_d^I + U^I$. This is a consequence of the fact that due to the normal ordering a factor $n(\epsilon)$ appears in (4.24). This result is generic and holds for a general function $J(\epsilon, 0)$. A similar effect occurs at the band edge if $J(\epsilon, 0)$ is not a continuous function at the band edge. Then we obtain singularities in the integral in (4.24) if $\epsilon_d^R$ or $\epsilon_d^R + U^R$ approach the band edge. Consequently these quantities behave discontinuously at the band edge as a function of the initial values.

There is another interesting point that we would like to mention. In the symmetric case $U = -2\epsilon_d$ the renormalized value of $U$ does not depend on the temperature. If the system deviates only a bit from the symmetric case, (4.24) and (4.25) show that the system is pushed in the direction of the symmetric case. This can be seen if one takes $\epsilon_d = -U/2 + \delta$ and expands the right hand side of (4.24). The renormalized value of $\delta$ is smaller than the initial value of $\delta$. In this sense the symmetric situation is stable. Near the symmetric point the temperature dependence of the renormalized values will be weak. But in the general case the renormalized values of the impurity energy and the interaction will depend on the temperature. For small enough temperature the integral in (4.24) can be evaluated using the usual Sommerfeld expansion. This yields

$$\epsilon_d^R = \epsilon_d^I + \int d\epsilon J(\epsilon, 0)\frac{\epsilon_d^R - \epsilon + (1 + \theta(-\epsilon))U^R}{(\epsilon_d^R - \epsilon)(\epsilon_d^R - \epsilon + U^R)} + \frac{\pi}{6}\Gamma(k_B T)^2\left((\epsilon_d^R)^{-2} - (\epsilon_d^R + U^R)^{-2}\right) + O(T^4) \tag{4.26}$$

This shows that for $\left|\epsilon_d^R + U^R\right| > \left|\epsilon_d^R\right|$ we obtain $\epsilon_d^R(T) > \epsilon_d^R(T = 0)$, whereas for $\left|\epsilon_d^R + U^R\right| < \left|\epsilon_d^R\right|$ we obtain $\epsilon_d^R(T) < \epsilon_d^R(T = 0)$. Generally Eq. (4.26) is a good approximation only for $T < \left|\epsilon_d^R\right|$:

When $T$ becomes larger one can show that $\epsilon_d^R(T)$ will decrease as a function of $T$ for $\left|\epsilon_d^R + U^R\right| > \left|\epsilon_d^R\right|$. The temperature dependence of $\epsilon_d^R$ leads to a weak temperature dependence of $U^R$. In the special case $J(\epsilon, 0) \propto \sqrt{D^2 - \epsilon^2}$ we obtain a temperature dependence of $U^R$ only if $\epsilon_d^R$ or $\epsilon_d^R + U^R$ lie outside the band. If $\left|\epsilon_d^R\right|, \left|\epsilon_d^R + U^R\right| < D$, (4.12) shows that $U^R$ does not depend on the temperature.

## 5 The induced spin-spin interaction

So far the contribution (2.8), which is generated by the unitary transformation, and which therefore has to be included in the Hamiltonian, was not taken into account. This interaction gives rise to a spin–spin coupling term

$$-2 \sum_{k,q} V_{k,q}^{(2)} \, (\psi_k^\dagger \frac{1}{2} \vec{\sigma} \, \psi_q) \cdot (\psi_d^\dagger \frac{1}{2} \vec{\sigma} \, \psi_d) \tag{5.1}$$

with

$$\psi_k = \begin{pmatrix} c_{k,+} \\ c_{k,-} \end{pmatrix}, \psi_d = \begin{pmatrix} d_+ \\ d_- \end{pmatrix}, \tag{5.2}$$

the so called potential scattering term

$$\frac{1}{2} \sum_{k,q} V_{k,q}^{(2)} \, (\psi_k^\dagger \psi_q) \, (\psi_d^\dagger \psi_d), \tag{5.3}$$

and a term

$$\frac{1}{2} \sum_{k,q} V_{k,q}^{(2)} \, (c_{k,\sigma}^\dagger c_{q,-\sigma}^\dagger d_{-\sigma} d_\sigma + d_\sigma^\dagger d_{-\sigma}^\dagger c_{q,-\sigma} c_{k,\sigma}) \, . \tag{5.4}$$

The final Hamiltonian contains no couplings between states that have a singly occupied impurity orbital and states for which the impurity orbital is either empty or doubly occupied. Whereas the spin–spin coupling (5.1) acts only on the part of the Hilbert space of states with a singly occupied impurity orbital, the term (5.4) vanishes on this part of the Hilbert space. It is important if the impurity orbital is either empty or doubly occupied. In this sense, these two terms are conjugate to each other. In fact one has a simple interpretation for these couplings in the symmetric Anderson model. Whereas the asymmetric Anderson model has only the usual SU(2)–spin symmetry, the symmetric Anderson model has an additional SU(2)–pseudo–spin symmetry. Introducing the wave vector $\pi$, which has all components equal to $\pi$, the symmetric energy band has the symmetry $\epsilon_k = -\epsilon_{\pi-k}$, $V_k = V_{\pi-k}$. For $\epsilon_d = -U/2$ the Hamiltonian also commutes with the operators

$$\begin{aligned}
\hat{S}_z &= \frac{1}{2} \left( 1 - d_+^\dagger d_+ - d_-^\dagger d_- + \sum_k (1 - c_{k,+}^\dagger c_{k,+} - c_{k,-}^\dagger c_{k,-}) \right) \\
\hat{S}_+ &= d_+ d_- + \sum_k c_{\pi-k,+} c_{k,-} \\
\hat{S}_- &= d_+^\dagger d_-^\dagger + \sum_k c_{k,-}^\dagger c_{\pi-k,+}^\dagger .
\end{aligned} \tag{5.5}$$

These operators form the second SU(2) symmetry mentioned above. The potential scattering term (5.3) and the term (5.4) together can be written as a pseudo–spin interaction. It is clear that if the original Hamiltonian has these symmetries, the transformed Hamiltonian has these symmetries too. Therefore the term (5.4) is present if the corresponding spin–spin interaction is present. Although in the asymmetric case the Hamiltonian does not have the additional symmetry, the two terms (5.3) and (5.4) have the same interpretation. The only difference is that now the coupling constant $V_{k,q}^{(2)}$ is not symmetric with respect to a transformation $k \to \pi - k$.

In the remaining part of this section we want to discuss the regime where the impurity orbital is singly occupied. Since $\psi_d^\dagger \psi_d = 1$ in this regime, the pseudo–spin interaction reduces to a scattering of band electrons. Such a term has already been neglected and therefore this contribution is not taken into account.

In spite of the fact that the additional couplings $V_{k,q}^{(2)}$ are small compared to the other parameters of the Anderson model, they cannot be ignored. In the regime where the impurity orbital is singly occupied, $\epsilon_d^R < \epsilon_F$ and $\epsilon_d^R + U^R > \epsilon_F$, even a small antiferromagnetic spin–spin coupling at the Fermi surface $V_{k_F,k_F}^{(2)} < 0$ gives rise to the Kondo effect for low temperatures. In the Kondo model the Kondo temperature can be defined as $T_K = (2\pi \chi_{\text{imp}}(T=0))^{-1}$ [4] where $\chi_{\text{imp}}(T=0)$ is the impurity contribution to the susceptibility at zero temperature. Let us remark that other definitions of the Kondo temperature can be found in the literature, the definition by Wilson [5] is somewhat different. For a detailed discussion see e.g. [4]. Based on a Bethe-ansatz solution, Tsvelick and Wiegmann argue that the Kondo temperature $T_K$ is given by

$$k_B T_K = \frac{2}{\pi} D \, \exp\left[-\Phi(2\rho(\epsilon_F) V_{k_F,k_F}^{(2)})\right] \tag{5.6}$$

with the universal function [5]

$$\Phi(y) = \frac{1}{|y|} - \frac{1}{2} \ln |y| + O(y). \tag{5.7}$$

For the Kondo problem $D$ is the conduction band width. In the Anderson impurity model one knows that $D$ in (5.6) has to be replaced by an *effective band width* $D_{\text{eff}}$ that cannot be larger than $U^R$ [10]. If one follows the perturbative calculation of e.g. the susceptibility in the Kondo problem, one notices that the breakdown of the perturbation expansion is due to the matrix elements $V_{k_F,q}^{(2)}$: These describe the scattering of an electron from the Fermi surface with the impurity to some wave vector $q$ and then back to the Fermi surface. For this reason one is not only interested in the coupling at the Fermi surface $V_{k_F,k_F}^{(2)}$, but also in $V_{k_F,q}^{(2)}$ since this determines the effective band width of the associated Kondo problem.

Let us now calculate the matrix elements $V_{k,q}^{(2)}$ in the flow equations approach. We already mentioned that the additional couplings (2.8) do not lead to a contribution in the equations for $\epsilon_d$ and $U$. Therefore we calculate the coupling constant simply by integrating the coefficient in front of the interaction term in (5.1). To be consistent with the notation in our previous paper we symmetrize this coefficient. We have to calculate

$$V_{k,q}^{(2)} = \int_0^\infty d\ell (\eta_k^{(2)} V_q + \eta_q^{(2)} V_k). \tag{5.8}$$

Using (2.7) and (3.3) to replace $\eta_k^{(2)}$ and furthermore (4.4), we obtain

$$V_{k,q}^{(2)} = -\frac{1}{2} \int_0^\infty d\ell V_k V_q U \left( \frac{\frac{\partial \ln J(\epsilon_k,\ell)}{\partial \ell}}{(\epsilon_d - \epsilon_k)(\epsilon_d - \epsilon_k + U)} + \frac{\frac{\partial \ln J(\epsilon_q,\ell)}{\partial \ell}}{(\epsilon_d - \epsilon_q)(\epsilon_d - \epsilon_q + U)} \right). \tag{5.9}$$

The $\ell$-dependence of $V_k$ is obtained from (4.1). Using again (3.3) and (4.4) we obtain

$$V_k(\ell) = V_k(0) \sqrt{\frac{J(\epsilon_k,\ell)}{J(\epsilon_k,0)}}. \tag{5.10}$$

This yields

$$V^{(2)}_{k,q} = -\frac{1}{2} V_k(0) V_q(0) \int_0^\infty d\ell\, U (J(\epsilon_k,\ell) J(\epsilon_k,0) J(\epsilon_q,\ell) J(\epsilon_q,0))^{-\frac{1}{2}}$$
$$\times \left( \frac{\frac{\partial J(\epsilon_k,\ell)}{\partial \ell} J(\epsilon_q,\ell)}{(\epsilon_d - \epsilon_k)(\epsilon_d - \epsilon_k + U)} + \frac{\frac{\partial J(\epsilon_q,\ell)}{\partial \ell} J(\epsilon_k,\ell)}{(\epsilon_d - \epsilon_q)(\epsilon_d - \epsilon_q + U)} \right). \tag{5.11}$$

Using the parametrization for $J(\epsilon,\ell)$ introduced in (4.7), this expression simplifies to

$$V^{(2)}_{k,q} = \frac{1}{2} V_k(0) V_q(0) \int_0^\infty d\ell\, U \frac{(\epsilon_d - \epsilon_k)(\epsilon_d - \epsilon_k + U) + (\epsilon_d - \epsilon_q)(\epsilon_d - \epsilon_q + U)}{\epsilon_d^2 + (\epsilon_d + U)^2}$$
$$\times \exp\left( -\frac{1}{2} \int_0^\ell \left( \frac{(\epsilon_d - \epsilon_k)^2(\epsilon_d - \epsilon_k + U)^2}{\epsilon_d^2 + (\epsilon_d + U)^2} + \frac{(\epsilon_d - \epsilon_q)^2(\epsilon_d - \epsilon_q + U)^2}{\epsilon_d^2 + (\epsilon_d + U)^2} \right) d\ell' \right). \tag{5.12}$$

Let us replace $\epsilon_d$ and $U$ on the right hand side by their renormalized values. The same reasoning applies with respect to this approximation as at the end of section 3, in particular for the important matrix elements at the Fermi surface. One finds

$$V^{(2)}_{k,q} = V_k(0) V_q(0) U^R$$
$$\times \frac{(\epsilon_d^R - \epsilon_k)(\epsilon_d^R - \epsilon_k + U^R) + (\epsilon_d^R - \epsilon_q)(\epsilon_d^R - \epsilon_q + U^R)}{(\epsilon_d^R - \epsilon_k)^2(\epsilon_d^R - \epsilon_k + U^R)^2 + (\epsilon_d^R - \epsilon_q)^2(\epsilon_d^R - \epsilon_q + U^R)^2}. \tag{5.13}$$

This formula is a very good approximation to (5.12) if $\epsilon_k$ or $\epsilon_q$ are not too close to $\epsilon_d^R$ or $\epsilon_d^R + U^R$. If both band energies become equal to $\epsilon_d^R$ or $\epsilon_d^R + U^R$, the approximate result diverges. In this special case the asymptotic behaviour of $\epsilon_d$ and $U$ becomes important. One can show that the integral in (5.12) has a logarithmic divergence, which yields a logarithmic divergence of $V^{(2)}_{k,q}$ if $\epsilon_k$ and $\epsilon_q$ approach $\epsilon_d^R$ or $\epsilon_d^R + U^R$. Such a divergence causes no problems since it is integrable as a function of the energy $\epsilon_k$; also higher powers of $V^{(2)}_{k,q}$ remain integrable.

When calculating the Kondo temperature, we only need $V^{(2)}_{k,q}$ for the case where at least one of the band energies is equal (or at least very close) to the Fermi energy. Since $\epsilon_d^R$ or $\epsilon_d^R + U^R$ can only be equal to the Fermi energy if the hybridization vanishes, we can use (5.13) in the following. At the Fermi surface this yields

$$V^{(2)}_{k_F,k_F} = V_{k_F}(0)^2 \frac{U^R}{\epsilon_d^R(\epsilon_d^R + U^R)}, \tag{5.14}$$

and with only one wave vector at the Fermi surface

$$V^{(2)}_{k_F,q} = V_{k_F}(0)\, V_q(0)\, U^R\, \frac{\epsilon_d^R(\epsilon_d^R + U^R) + (\epsilon_d^R - \epsilon_q)(\epsilon_d^R - \epsilon_q + U^R)}{(\epsilon_d^R)^2(\epsilon_d^R + U^R)^2 + (\epsilon_d^R - \epsilon_q)^2(\epsilon_d^R - \epsilon_q + U^R)^2}. \tag{5.15}$$

Before proceeding with the calculation of the Kondo temperature in various cases, it is interesting to compare this result with the coupling obtained by the Schrieffer–Wolff unitary transformation in Ref. [9]. There one finds

$$V^{(2)}_{k,q} = \frac{1}{2} V_k(0)\, V_q(0)\, U^I \left( \frac{1}{(\epsilon_d^I - \epsilon_k)(\epsilon_d^I - \epsilon_k + U^I)} + \frac{1}{(\epsilon_d^I - \epsilon_q)(\epsilon_d^I - \epsilon_q + U^I)} \right), \tag{5.16}$$

in particular at the Fermi surface

$$V^{(2)}_{k_F,k_F} = V_{k_F}(0)^2 \frac{U^I}{\epsilon_d^I(\epsilon_d^I + U^I)}. \tag{5.17}$$

As a first remark we mention that for $\epsilon_k = \epsilon_q$ both results are identical if one replaces the initial values of $\epsilon_d^I$ and $U^I$ in the result by Schrieffer and Wolff with the renormalized values $\epsilon_d^R$ and $U^R$. For $\epsilon_k \neq \epsilon_q$ our result differs from the Schrieffer–Wolff result.

The first problem with the induced spin–spin interaction in the formalism of Schrieffer and Wolff are the pole terms in $V_{k_F,q}^{(2)}$ if $\epsilon_d^I$ or $\epsilon_d^I + U^I$ lie in the conduction band. A second problem is apparent in the following limit

$$V_{k_F,q}^{(2)} \stackrel{|\epsilon_q| \to \infty}{\longrightarrow} \frac{1}{2} V_{k_F}(0) V_q(0) \frac{U^I}{\epsilon_d^I(\epsilon_d^I + U^I)} \neq 0. \tag{5.18}$$

This immediately implies that the effective band width in the corresponding Kondo problem is of order the conduction band width

$$D_{\text{eff}} \propto D, \tag{5.19}$$

which is known to be wrong. It is quite obvious from Eq. (5.15) that both these problems do not show up in the flow equations approach.

In order to obtain some more quantitative results, let us now discuss two particular regimes of the Anderson model.

### Symmetric Anderson model with $\left|\epsilon_d^R\right| < D$

In the symmetric case one has $\epsilon_d^R = -U^R/2$ and the relevant matrix elements of the spin–spin coupling are

$$V_{k_F,q}^{(2)} = V_{k_F}(0) V_q(0) U^R \frac{\epsilon_q^2 - \frac{(U^R)^2}{2}}{\left(\epsilon_q^2 - \frac{(U^R)^2}{4}\right)^2 + \left(\frac{(U^R)^2}{4}\right)^2}. \tag{5.20}$$

This is depicted in Fig. 3 where it can be compared with the Schrieffer–Wolff result. For simplicity we have assumed $V_q(0) = V_{k_F}(0)$ for all wave vectors $q$ in the diagram. Furthermore, we have replaced $U^I$ by $U^R$ in the Schrieffer–Wolff result, see also our discussion in section 6 for this point. The Kondo temperature depends mainly on the coupling at the Fermi surface (compare Eq. (5.6))

$$2\rho(\epsilon_F) V_{k_F,k_F}^{(2)} = -\frac{8\Gamma}{\pi U^I}\left(1 - \frac{2\Gamma}{D}\right). \tag{5.21}$$

This is consistent with the Schrieffer–Wolff result except that we find an additional (usually small) correction term $2\Gamma/D$.

The main difference to the Schrieffer–Wolff result shows up in the effective band width of the associated Kondo problem. In the flow equations approach the effective band width is obviously proportional to the on–site interaction

$$D_{\text{eff}} \propto U^R \tag{5.22}$$

since the spin–spin coupling becomes ferromagnetic for $|\epsilon_q| > U^R/\sqrt{2}$ and decays to zero even further away from the Fermi surface. Since the spin–spin coupling induced by our unitary transformation is not constant (there is no physical reason why it should be), it is difficult to say quantitatively what the proportionality factor in Eq. (5.22) is: The Kondo problem is usually only treated for a constant spin–spin coupling with the conduction band electrons. Nevertheless, we would like to estimate the proportionality constant in (5.22) for the special case of a constant density of states and constant $V_q(0)$. This value can be compared with known results. To get some rather approximate estimate one can e.g. replace $V^{(2)}_{k_F,q}$ regarded as a function of $\epsilon_q$ by its value at the Fermi surface in an interval around the Fermi surface that has the same area as the original curve. This should give a lower bound on $D_{\text{eff}}$. We obtain

$$D_{\text{eff}} \gtrsim 0.36\, U^R. \tag{5.23}$$

This result can be compared with the result from the Bethe-ansatz solution. There one obtains

$$k_B T_K = \frac{U}{\sqrt{4\pi}} \exp(-\Phi(2\rho(\epsilon_F)V^{(2)}_{k_F,k_F})). \tag{5.24}$$

We identify the prefactor with the one in (5.6) and obtain

$$D_{\text{eff}} = \frac{\sqrt{\pi}}{4} U = 0.443\, U. \tag{5.25}$$

This is in good agreement with (5.23). Estimates similar to (5.23) can be obtained for an arbitrary density of states and arbitrary $V_q(0)$.

### Valence fluctuation regime $0 < -\epsilon_d^R \ll U^R \ll D$

This regime shows characteristic new features of the asymmetric Anderson model. The spin–spin coupling is given by

$$V^{(2)}_{k_F,q} = V_{k_F}(0)\, V_q(0)\, \frac{2\epsilon_d^R - \epsilon_q}{(\epsilon_d^R - \epsilon_q)^2 + (\epsilon_d^R)^2} \tag{5.26}$$

with the value at the Fermi surface

$$2\rho(\epsilon_F)\, V^{(2)}_{k_F,k_F} = \frac{2\Gamma}{\pi \epsilon_d^R}. \tag{5.27}$$

Essentially the same remarks apply as in the previous section. The coupling is depicted in Fig. 4 where it can be compared with the Schrieffer–Wolff result (again with $V_q(0) = V_{k_F}(0)$ and with $\epsilon_d^I$, $U^I$ replaced by $\epsilon_d^R$, $U^R$). As before the flow equations approach yields the correct scaling behaviour of the effective band width

$$D_{\text{eff}} \propto \left|\epsilon_d^R\right|. \tag{5.28}$$

It is important that the "renormalized" value of the impurity orbital energy enters in Eqs. (5.26) and (5.27). This behaviour is known from perturbative renormalization [10, 16] or numerical renormalization [6] which give the same value of the renormalized impurity orbital energy that we have calculated in Eq. (4.21). In this regime the Schrieffer–Wolff unitary transformation not only gives the wrong scaling behaviour of the effective band width $D_{\text{eff}} \propto D$, but also the wrong coupling at the Fermi surface since the initial value of the impurity orbital energy enters.

At this point one can wonder about the temperature dependence of $\epsilon_d^R$ that has been found in Eq. (4.26). This effect seems to be unobserved in renormalization treatments. However, the maximum effect of non–zero temperature is to increase $\epsilon_d^R(T=0)$ by a value of order $\Gamma$. According to Fig. 2 the smallest value of $\left|\epsilon_d^R\right|$ with $\epsilon_d^R < 0$ is $\epsilon_d^R = -\Gamma/\pi$, that is of order $\Gamma$. Thus one might expect to see some influence of the temperature dependence of $\epsilon_d^R$ in this regime with $\left|\epsilon_d^R\right| \lesssim \Gamma$. But this is just the mixed valence regime (for $k_B T \lesssim \Gamma$) where scaling breaks down anyway [16]. For this reason there is no contradiction.

# 6 Comparison with other methods I (Schrieffer–Wolff unitary transformation)

At this point we would like to explain in more detail why the Schrieffer–Wolff transformation and our continuous unitary transformation yield different results. At a first glance these two transformations are very similar. Our $\eta$ has the same general structure as the generator $S$ in the Schrieffer–Wolff transformation $H \to e^S H e^{-S}$ [9] with

$$S = \sum_{k,\sigma} V_k(0) \left( \frac{U^I}{(\epsilon_d^I - \epsilon_k)(\epsilon_d^I - \epsilon_k + U^I)} d^\dagger_{-\sigma} d_{-\sigma} c^\dagger_{k,\sigma} d_\sigma + \frac{1}{\epsilon_k - \epsilon_d^I} c^\dagger_{k,\sigma} d_\sigma \right) - h.c. \tag{6.1}$$

In both approaches the same more complicated interactions generated due to higher commutators are neglected. But as we have seen the two transformations still show important differences that have to be explained.

First of all, let us mention that it is possible to construct a continuous unitary transformation that reproduces the Schrieffer–Wolff result (5.16) for the coupling $V^{(2)}_{k,q}$ with the initial values of $\epsilon_d$ and $U$ replaced by their renormalized values. In fact one can achieve this by choosing the parametrization

$$f(\epsilon, \ell) = -\frac{1}{2} \frac{1}{\epsilon_d - \epsilon} \tag{6.2}$$

instead of Eq. (4.8). Then the hybridization $J(\epsilon, \ell)$ shows a very simple flow

$$J(\epsilon, \ell) = J(\epsilon, 0) \exp(-l). \tag{6.3}$$

This is depicted in Fig. 5 where it can be compared with the behaviour for our choice of $f(\epsilon, \ell)$ from Eq. (4.8). As compared to the original Schrieffer–Wolff transformation this is an improvement since now the renormalized parameters enter into the expressions for the induced spin–spin interaction at the Fermi surface $V^{(2)}_{k_F, k_F}$. In the valence fluctuation regime this is of importance as discussed in the previous section.

Still the main differences between the Schrieffer–Wolff result and the flow equations are not resolved so easily. That is the couplings $V^{(2)}_{k,q}$ show the wrong high–energy cutoff and contain pole terms. Obviously our result for the spin–spin coupling differs from the result by Schrieffer and Wolff already in second order in the hybridization $V_k$. This is due to the fact that our transformation differs in this order from the Schrieffer–Wolff transformation. In principle it is of course possible to write our transformation in the form $\exp(S)$ too. $S$ can be calculated from the expansion

$$S = \int_0^\infty d\ell \, \eta(\ell) + \frac{1}{2} \int_0^\infty d\ell \int_0^\ell d\ell' [\eta(\ell), \eta(\ell')] + \ldots. \tag{6.4}$$

In second order in $V_k$ the term containing the commutator of $\eta$ at two different values of the flow parameter becomes important. It is of the form

$$\sum_{k,q,\sigma} S^{(2)}_{k,q} (: c^\dagger_{k,\sigma} d^\dagger_{-\sigma} d_{-\sigma} c_{q,\sigma} : - : c^\dagger_{q,\sigma} d^\dagger_{-\sigma} d_{-\sigma} c_{k,\sigma} : - : c^\dagger_{k,\sigma} d^\dagger_{-\sigma} d_\sigma c_{q,-\sigma} :$$
$$+ : c^\dagger_{q,\sigma} d^\dagger_{-\sigma} d_\sigma c_{k,-\sigma} : + c^\dagger_{k,\sigma} c^\dagger_{q,-\sigma} d_{-\sigma} d_\sigma - d^\dagger_\sigma d^\dagger_{-\sigma} c_{q,-\sigma} c_{k,\sigma}). \tag{6.5}$$

Our choice of $f(\epsilon, \ell)$ leads to a controlled expansion without any pole terms here: The worst divergencies occuring in the generated couplings are only logarithmic pole terms and are therefore themselves integrable (compare the discussion in section 5). In contrast the parametrization (6.2) leads to non–integrable pole terms in (6.4) that cannot be interpreted as principal value integrals.

Of course terms in order $V_k^2$ curing the divergencies could be introduced by hand into the original Schrieffer–Wolff generator $S$ in Eq. (6.1). However, these would be divergent too, difficult to construct, and difficult to control in higher orders. In the flow equations formalism such terms appear naturally since $\eta$ does not commute with itself for different values of $\ell$. Also the problems with pole terms are (as far as we have seen) resolved due to the introduction of a flow parameter that generates an asymptotic behaviour of the parameters and thereby "smears out" the pole terms.

## 7   Comparison with other methods II (Anderson's "poor man's" scaling)

Let us compare our method to perturbative renormalization in the spirit of Anderson's "poor man's" scaling approach [15]. We will briefly review the main features of this approach as applied to the Anderson impurity model by Haldane [16]. For simplicity we assume $U \gg D \gg |\epsilon_d|$. The band width $D$ is reduced to $D - dD$ by perturbatively integrating out states with energies $D - dD < |\epsilon| < D$. This results in the following scaling equations [16]

$$\frac{d\epsilon_d}{d \ln D} = -\int_0^\infty d\epsilon \left( \frac{2\epsilon \frac{\partial}{\partial \epsilon} J^I(-\epsilon)}{\epsilon + \epsilon_d} + \frac{\epsilon \frac{\partial}{\partial \epsilon} J^I(\epsilon)}{\epsilon - \epsilon_d} \right) \tag{7.1}$$

$$\frac{d\Gamma}{d \ln D} = O(\Gamma J^I(D)/D). \tag{7.2}$$

In order to avoid confusion with the flow equations approach we have introduced the notation $J^I(\epsilon) \stackrel{\text{def}}{=} J(\epsilon, \ell = 0)$. The second equation leads to no nontrivial scaling behaviour. In the first equation the main contributions come from $|\epsilon| \approx D$ and Haldane finds

$$\frac{d\epsilon_d}{d \ln D} = -\frac{\Gamma}{\pi} + O(\epsilon_d J^I(D)/D). \tag{7.3}$$

By scaling down to $D \approx k_B T$ one obtains the effective impurity orbital energy $E_d^*$ that is equal to our $\epsilon_d^R$ in (4.19). In the limit $D \gg U \gg \left|\epsilon_d^R\right|$ one argues that non-trivial scaling occurs only when $D$ has been reduced to order $U$. In this case one obtains (4.22) for the effective impurity energy that enters into the induced spin–spin interaction.

Now the scaling equation (7.1) can also be obtained in the flow equations formalism. One has to choose

$$J(\epsilon, \ell) \stackrel{\text{def}}{=} J^I \left( \epsilon \frac{D^I}{D(\ell)} \right), \tag{7.4}$$

where $D^I$ is the initial band width. $D(\ell)$ is a parameter (effective band width) that approaches zero monotonously as $\ell \to \infty$. Instead of using the unfamiliar parameter $\ell$ we can then try to work with the more familiar parameter $D$. Obviously (7.4) amounts to decoupling the high–energy modes by means of a unitary transformation as depicted in Fig. 5. It should come as no surprise that this yields flow equations in $D$ reminiscent of the familiar scaling equations. Eq. (7.4) corresponds to a certain choice of $f(\epsilon, \ell)$ but this will be of no importance for our discussion.

The flow equation (4.5) for $\epsilon_d$ in the limit $U \to \infty$ expressed as a function of $D = D(\ell)$ takes the following form after a short calculation

$$\frac{d\epsilon_d}{d \ln D} = \int d\epsilon (1 + n(\epsilon)) \frac{\epsilon}{\epsilon_d - \epsilon} \frac{\partial}{\partial \epsilon} J^I(\epsilon \frac{D^I}{D}) \tag{7.5}$$

to be integrated from $D = D^I$ to $D = 0$. For $T = 0$ this obviously just reproduces Haldane's scaling equation (7.1).

Now first of all this might make some approximations more plausible that we have used in the flow equations approach. In particular the restriction to a certain set of interactions is standard in a renormalization framework. It is well–known that the procedure of integrating out high–energy modes leads to additional interactions in the "poor man's" approach [16]. This will also include e.g. the induced spin–spin interaction. However, even these terms are not taken into account in the scaling approach.[1] Instead one argues that on a low–energy scale the physical behaviour is determined by the scaling invariants like $E_d^*$.

Though intuitively quite appealing, the parametrization (7.4) leads to immediate problems in the flow equations formalism since we do not use similar approximations. The first problem is already apparent from (7.2): Why should the infinitesimal unitary transformations stop when $D$ is of order $k_B T$? Such a cutoff is necessary, otherwise Eq. (7.2) leads to divergent behaviour as $D \to 0$.

An even worse problem shows up when comparing the self–consistency equations (4.24) and (4.25) with the actual solution of the differential equations in $D$. Let us for simplicity consider the symmetric Anderson model with (compare Eq. (4.13))

$$U^R = \frac{U^I}{1 - \frac{2\Gamma}{D}} < \frac{D}{2}. \tag{7.6}$$

However, the solution of the differential equation in $D$ leads to $U(\ell = \infty) = 0$ as can be shown rigorously for the semi–circle case. Intuitively what happens is that the modes with energies larger than $U/2$ that are being integrated out push $U$ closer to the Fermi surface. Before the low–lying modes can push $U$ away again, one has reached $U = 0$. Just this behaviour is avoided by the parametrization (4.7), compare also Fig. 5. This is the reason why the decoupling in our case can be performed all the way down to $\ell = \infty$ without introducing a condition like $D \approx k_B T$. The self–consistency conditions (4.24) and (4.25) have been obtained from (4.10) and (4.11) by taking $\ell = 0$. Thus the self–consistency conditions fail if the principal value integrals in (4.10) and (4.11) are not well–defined for some value of $\ell$. This happens with the parametrization (7.4). For $\ell$ such that $D(\ell) = \epsilon_d^R$ or $D(\ell) = \epsilon_d^R + U^R$ the integrals diverge, a familiar band–edge problem.

The reason why this phenomenon is unobserved in the scaling approach is that it is due to neglected terms like on the right hand side of Eq. (7.2). But if one attempts to study the behaviour when $U$ becomes of order $D$, these terms cannot be neglected. Another way of saying this is that the scaling equations do not reproduce a standard perturbational expansion like e.g. Eq. (7.6) and a behaviour like Eq. (4.19) within one framework. The flow equations manage this, but not with a parametrization like (7.4). One might hope that they can provide a tool to study crossover phenomena.

---

[1] In this sense our approximations are clearly better though less familiar.

# 8 Conclusions

We have shown that using a continuous unitary transformation it is possible to eliminate the hybridization term in the single impurity Anderson model. Similar to the well–known Schrieffer–Wolff transformation, an additional spin–spin interaction and pseudo–spin interaction are generated. Furthermore the impurity energy and the on–site interaction are renormalized. The main approximation was to neglect additional interactions generated by the transformation if the spin–spin interaction is included in the Hamiltonian. Instead we simply integrated the coefficient of this interaction to obtain the new coupling constants. Some justification for this procedure is that by introducing the spin–spin interaction in the Hamiltonian right from the beginning, we do not obtain additional contributions to the impurity energy and the interaction $U$. Such contributions occur only if even more complicated additional interactions are taken into account. In this manner we have restricted ourselves to some minimal but "consistent" set of interactions.

This procedure yields coupled non-linear flow equations for the impurity energy and the interaction. It turns out that these equations can be approximately solved by self–consistency conditions for the renormalized values of the impurity energy and the interaction. This approximation can be controlled by comparing the result with a numerical integration of the flow equations and it turns out that the approximation is very good. It becomes exact in the limit of vanishing interaction $U$. We discussed several regimes in the parameter space for a special coupling $J(\epsilon, 0) = \frac{\Gamma}{\pi D}\sqrt{D^2 - \epsilon^2}$. We do not expect that the general features depend on the details of $J(\epsilon, 0)$, but only on $\Gamma$ and $D$. The method works in the whole parameter space as long as the hybridization is not the largest energy scale in the system. Although we mainly discussed the case where $J(\epsilon, 0)$ is a semi–circle, the method can be applied to any other $J(\epsilon, 0)$ as well.

The Kondo regime of the Anderson impurity model has been investigated by mapping it onto the Kondo Hamiltonian with a unitary transformation. *In contrast to the Schrieffer–Wolff unitary transformation our approach of infinitesimal unitary transformations generates all the correct parameters of this Kondo Hamiltonian. In particular no pole terms appear in the spin–spin coupling. An important point is also that the correct high–energy cutoff enters in all the expressions for the renormalized or induced interactions.* One should expect that such a unitary transformation exists since after all the Anderson impurity model has been introduced to study dilute magnetic alloys which exhibit the Kondo effect.

Quite generally the flow equations seem to provide a good framework to investigate such induced interactions like the spin–spin interaction here. Another interesting problem of this type is the problem of electrons interacting with phonons in a solid. In this case the well-known Fröhlich transformation decouples the electrons from the phonons [17]. It generates an additional effective interaction between the electrons. In some region of the parameter space this interaction is attractive and is responsible for the formation of Cooper pairs. But again the Fröhlich transformation is ill–defined near resonances. Divergencies as in the case of the Schrieffer–Wolff transformation occur. This problem has recently been studied by Lenz [18] using continuous unitary transformations. As in our case his transformation is well–defined and the attractive interaction between the electrons does not show the usual divergencies.

The examples discussed so far show that the method of continuous unitary transformation has a wide range of applications. They provide a useful tool to simplify a given Hamiltonian. In the examples discussed so far [1, 7, 8, 18] a given Hamiltonian was transformed into a block–diagonal form by eliminating some of the coupling constants. This generates additional couplings and the original problem is mapped thereby to a different problem which can be investigated further. The main problem is that one has to neglect some of the additional couplings to obtain a closed set of equations. Up to now we are not able to present a general method that allows to estimate the effect of this approximation. The reasoning should probably be similar to the discussion of irrelevant terms in renormalization theory. But the comparison of our results with well known results obtained by other methods show that our approximation works well. In particular we can treat a strongly correlated system like the Anderson impurity model, where the natural small parameter is the hybridization $\Gamma$ and the rest of the Hamiltonian is not of simple quadratic form but contains the large interaction $U$.

A partially open problem is to calculate expectation values. To do this one has to apply the continuous unitary transformation to the observable as well. If one is only interested in static properties, this seems to be no problem. One has to solve the flow equations for the observable, which needs additional approximations for the observable similar to the approximations made for the Hamiltonian itself. Since we neglect only normal ordered terms, which do not contribute to the expectation value, this should be possible [1]. But if one wants to calculate dynamical properties like time–dependent correlation functions, it is yet not clear whether the same approximations can be applied. This problem has been discussed to some extent in Ref. [8].

One situation where one needs to calculate dynamical properties is the Hubbard model in the limit of infinite space dimensions [19]. It can be mapped onto a single impurity Anderson model with an additional self–consistency condition for the local density of states [20]. To apply our method to this problem, we have to calculate the single–particle Green's function, which is a dynamical problem. Also it would be important to attempt to study the crossover behaviour in the Kondo problem for low temperatures using our approach. Work in these directions is in progress. Other interesting problems are for example the single impurity Anderson models with two or more channels.

**Acknowledgement**

The authors would like to thank Prof. F. Wegner for many helpful discussions and suggestions. Furthermore we thank P. Lenz for discussions on related subjects. This work was supported in part by the Erwin Schrödinger Institute and by the Deutsche Forschungsgemeinschaft.

**Figure captions**

Fig. 1. Various regimes for the solution of the $U=0$ Anderson impurity model.

Fig. 2. The behaviour of the renormalized impurity orbital energy close to the Fermi surface as obtained from Eq. (4.19). The dashed line corresponds to the solution of the self–consistency condition, the full line is deduced from the numerical solution of the differential equations.

Fig. 3. Induced spin–spin interaction in the symmetric case.

Fig. 4. Induced spin–spin interaction in the $U \to \infty$ limit.

Fig. 5. Sketch of $J(\epsilon,\ell)$ for the symmetric Anderson impurity model for different values of $\ell$, $\ell_2 > \ell_1 > \ell_0 = 0$ or $D_2 < D_1 < D_0 = D^I$. $J(\epsilon,0)$ is the semi–circle.

In a) $J(\epsilon,\ell)$ is shown for the improved Schrieffer–Wolff transformation introduced in (6.3).

b) corresponds to the poor man's scaling approach in the context of the flow equations, compare sect. 7.

In c) the parametrization of $J(\epsilon,\ell)$ as chosen in this paper is depicted. The flow of $U(\ell)$ has been neglected in the diagram, in fact this would lead to an algebraic decay of $J(\epsilon,\ell)$ also for $\epsilon = \pm U^R/2$.

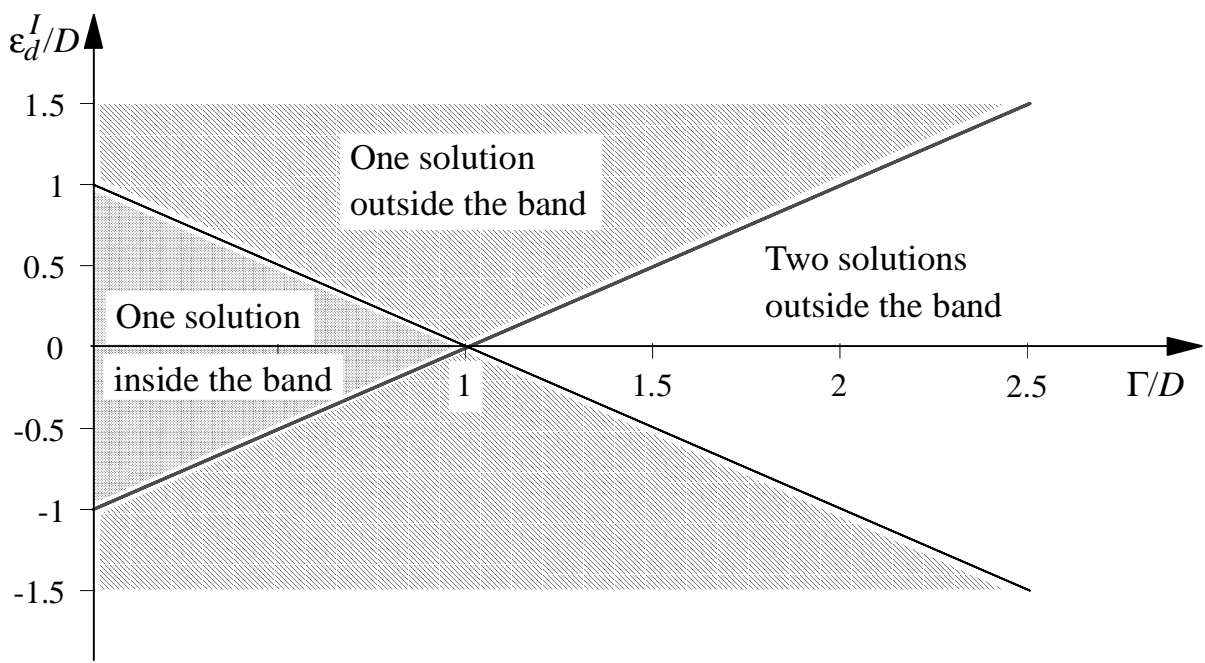

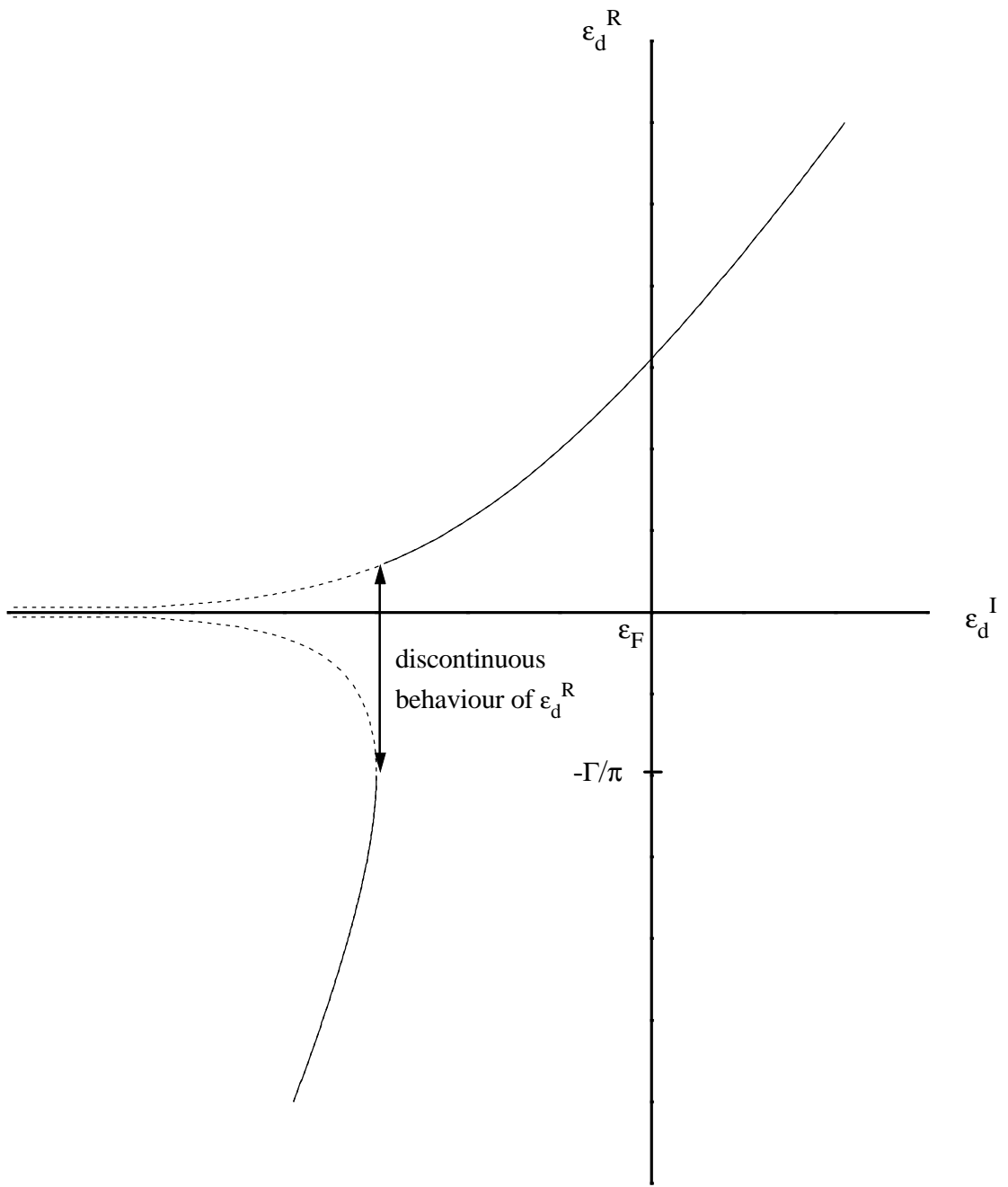

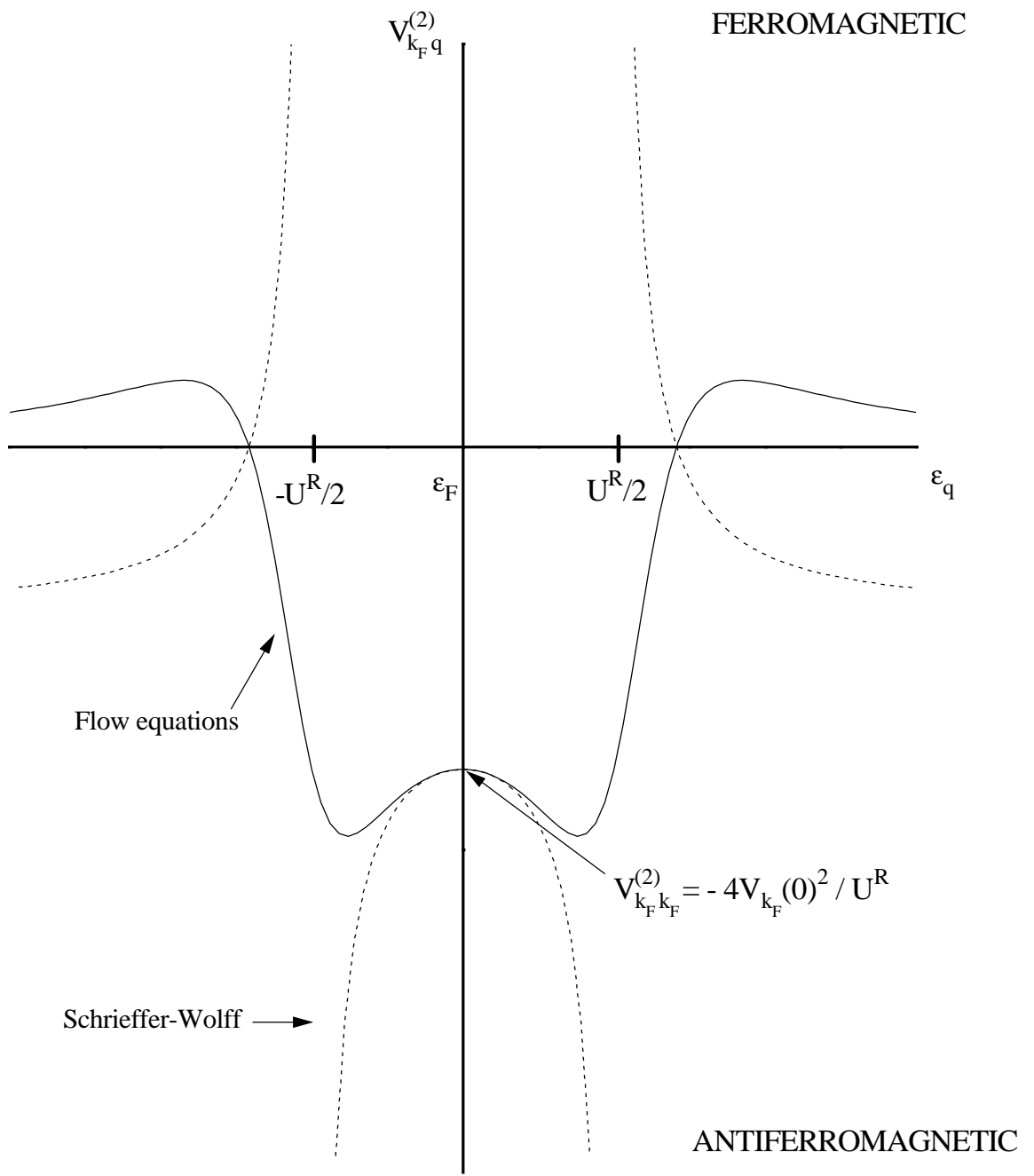

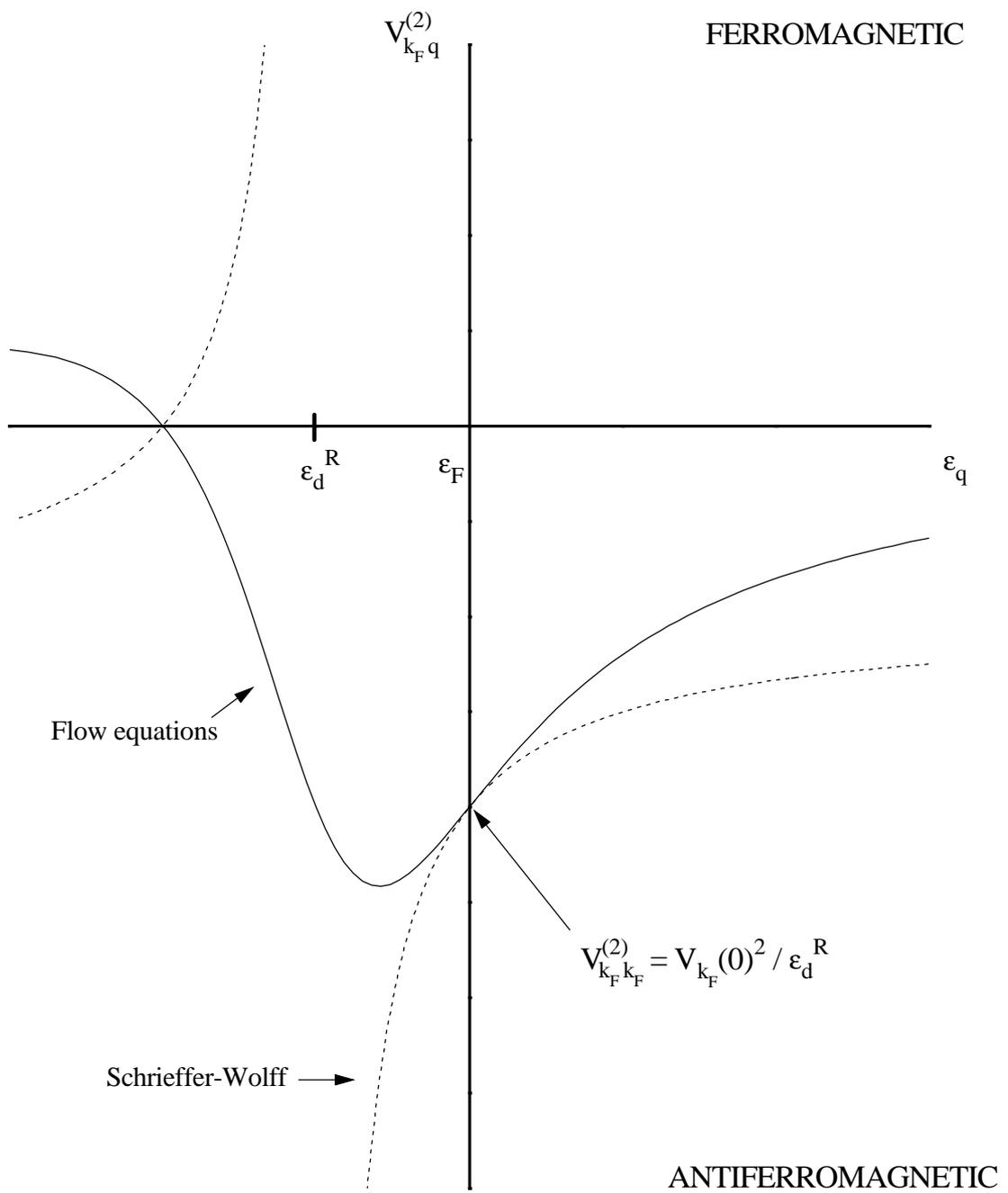

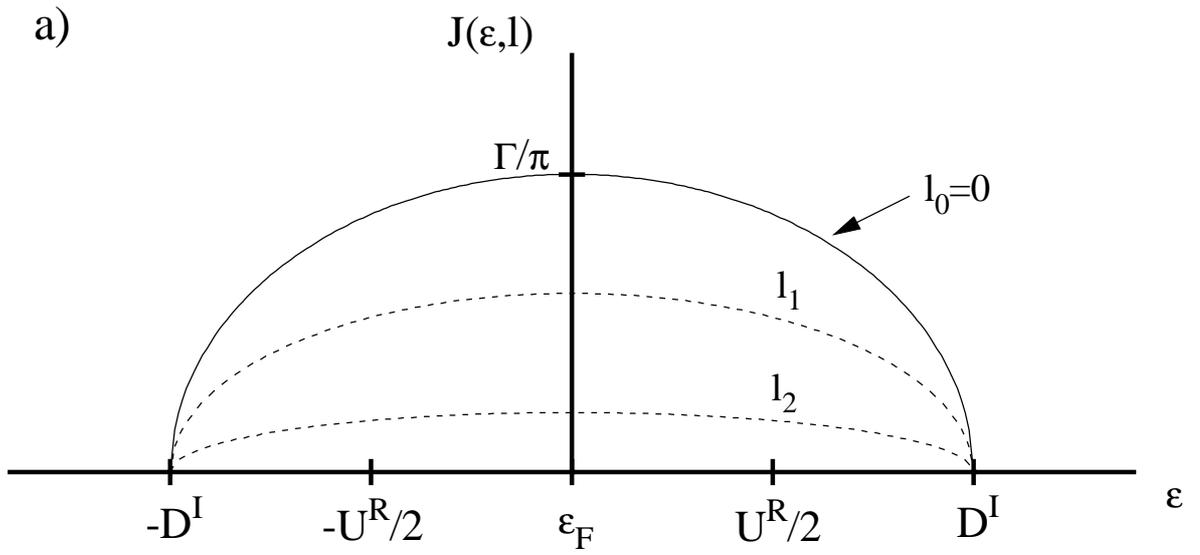

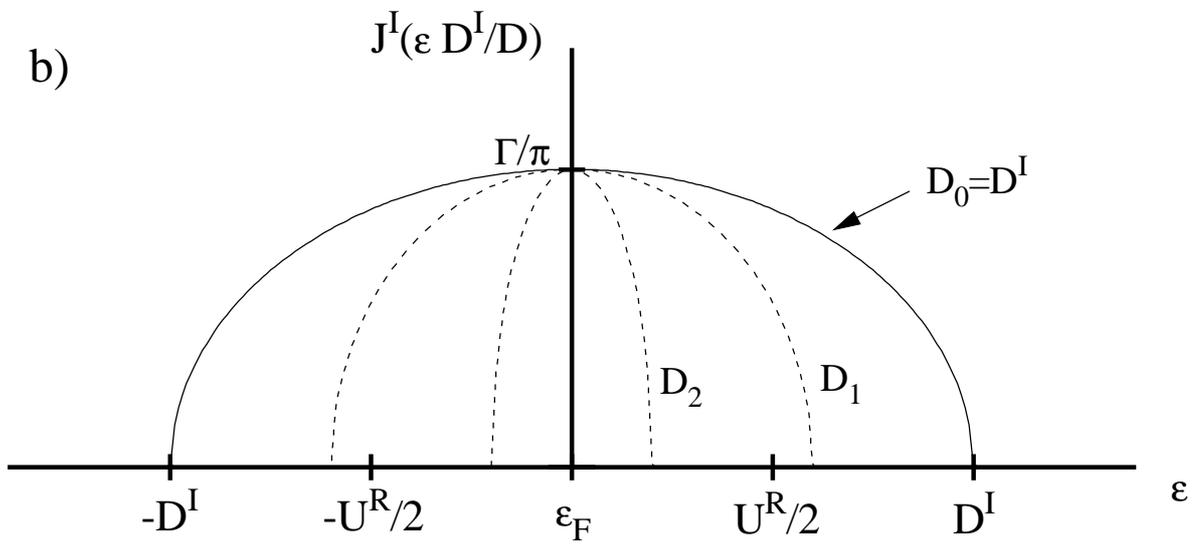

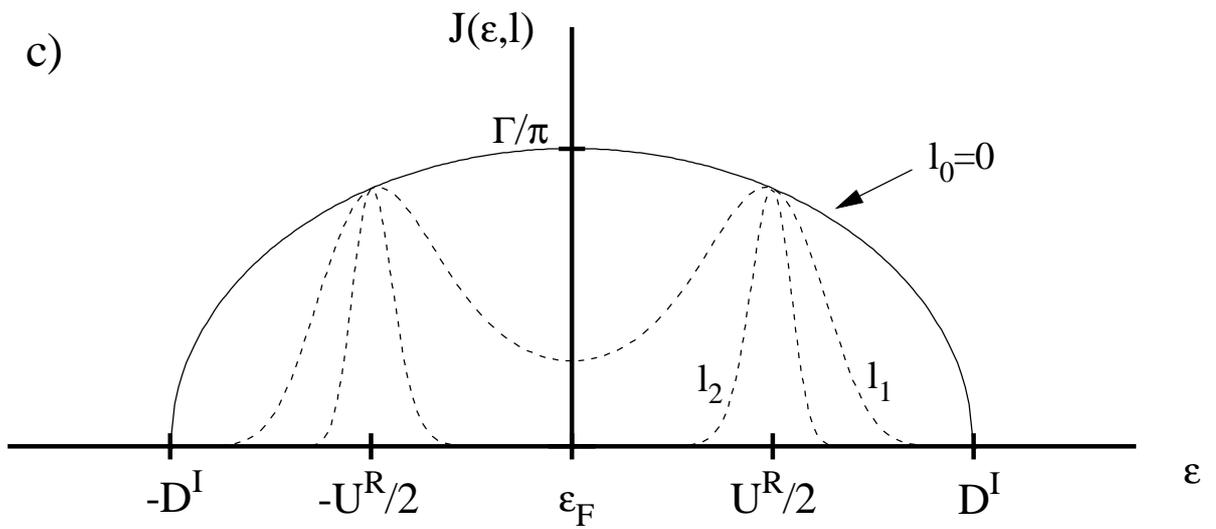